\documentclass{jfm}
\usepackage{amsmath,bm}
\usepackage{graphicx}
\usepackage{newtxtext}
\usepackage{newtxmath}
\usepackage{natbib}

\usepackage{caption}
\usepackage{subcaption}

\usepackage{hyperref}
\usepackage{placeins}
\hypersetup{
    colorlinks = true,
    urlcolor   = blue,
    citecolor  = black,
}

\newcommand{\be}{\begin{equation}}
\newcommand{\ee}{\end{equation}}
\newcommand{\bea}{\begin{eqnarray}}
\newcommand{\eea}{\end{eqnarray}}
\newcommand{\lb}{\label}

\newcommand{\RomanNumeralCaps}[1]
\linenumbers

\title{Lighthill's mechanism and vorticity cascade in the logarithmic layer of wall turbulence}
\author{Samvit Kumar\aff{1}
  \corresp{\email{skumar67@jh.edu}},
  Simon Toedtli \aff{2}, Tamer A.~Zaki \aff{1,2}
 \and Gregory Eyink\aff{1,2}}
\affiliation{\aff{1}Department of Applied Mathematics \& Statistics, Johns Hopkins University, Baltimore, MD, USA, 21218
\aff{2}Department of Mechanical Engineering, Johns Hopkins University, Baltimore, MD, USA, 21218}

\begin{document}
\maketitle

\begin{abstract}
We investigate Lighthill's proposed turbulent mechanism for near-wall concentration of spanwise vorticity by calculating mean flows conditioned on motion away from or toward the wall in an $Re_\tau=1000$ database of plane-parallel channel flow. Our results corroborate Lighthill's proposal throughout the entire logarithmic layer, but extended by counterflows that help explain anti-correlation of vorticity transport by advection and by stretching/tilting. We present evidence also for Lighthill's hypothesis that the vorticity transport in the log-layer is a ``cascade process'' through a scale-hierarchy
of eddies, with intense competition between transport outward from and inward to the wall. Townsend's model of attached eddies of hairpin-vortex type accounts for half of the vorticity cascade, whereas we identify 
necklace-type or  ``shawl
vortices'' that envelop turbulent sweeps as supplying the other half. 
%whereas the other half is carried by wide necklace-type vortices enveloping turbulent downflows, or sweeps. 
\end{abstract}

\section{Introduction}

In a famous review of boundary layers, 
\cite{lighthill1963} suggested that turbulent flows possess a mechanism that 
systematically concentrates spanwise vorticity sharply against the wall, despite the strong ``eddy viscosity''
effects that would be expected to diffuse such vorticity outward. He proposed that a tight correlation 
should exist, on the one hand, between motion toward the wall and vortex stretching/strengthening, and, on the other hand, between motion away from the wall and vortex compression/weakening (see \cite{lighthill1963}, section 3.3). Moreover, he argued that this 
mechanism should operate across the entire logarithmic layer and the correlated 
motions should constitute a turbulent ``cascade process'': 

\vspace{5pt}  
\noindent 
{\small ``We may think of them as constantly bringing the major part of the vorticity in the layer close to the wall, while intensifying it by stretching and, doubtless, generating new vorticity at the surface; meanwhile, they relax the vortex lines which they permit to wander into the outer layer. Smaller-scale movements take over from these to bring vorticity still closer to the wall, and so on. 
Thus, ... this cascade process has the additional effect in a turbulent boundary layer of bringing the fluctuations into closer and closer contact with the wall, while their vortex lines are more and more stretched.'' $\textendash$ \cite{lighthill1963}, p.99.}
\vspace{5pt}

\begin{figure}
%\hspace{60pt} 
%\centering
\includegraphics[width=0.9\textwidth]{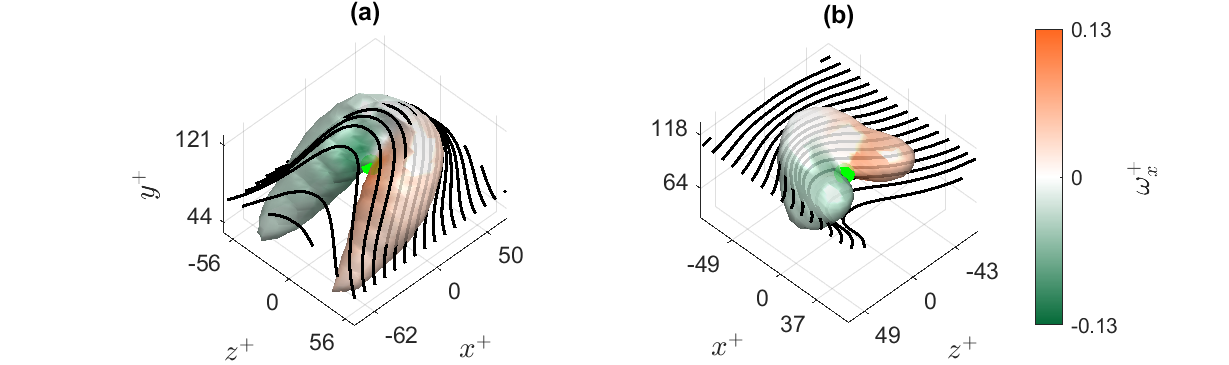}
\caption{Conditional eddies visualized for $\lambda_2=-0.9$, colored by $\omega_x^+$, 
along with vortex lines initiated at $y^+=60$ for (a) $v(y_c)>v_{rms}$ (\href{ https://cocalc.com/share/public_paths/cdee9e3cf9ce3d6804f9ed19f858a04b230f4250/fig1/eddy_streamline_outflow_yplus_093.html }{click for 3D version}) and at $y^+=108$ for (b) $v(y_c)<-v_{rms}$, 
(\href{ https://cocalc.com/share/public_paths/cdee9e3cf9ce3d6804f9ed19f858a04b230f4250/fig1/eddy_streamline_inflow_yplus_093.html }{click for 3D version}) with both conditions applied at $
y_c^+=92.8$. A green dot marks the conditioning point.}
\label{eddy}
\end{figure}

\noindent 
These ideas seem to have much in common with the attached eddy model (AEM) 
of \cite{townsend1976structure}, a popular vortex-based structure model of turbulent boundary layers which has been substantially further investigated and developed 
\citep{woodcock2015statistical,marusic2019attached}. 
In this approach, the boundary layer is modeled as a scale-invariant hierarchy of ``attached eddies'' often taken to be hairpin vortices similar to the structure visualized in Fig.~\ref{eddy}(a). These eddies are assumed to have dimensions which scale with wall distance $y$ and with a population size decreasing $\propto 1/y,$ consistent with an inverse cascade in which hairpin vortices generated by a bursting process lift from the wall, grow in size, and sequentially merge together. 

%{\sc \color{orange} xxxxxxxxxxxxxx The expression $\langle v\omega_z-w\omega_y\rangle$ has not been introduced, and $\Sigma$ is defined later in the text, in \S2.1. xxxxxxxxxxxxxx}\\
There have been sporadic attempts to unify Lighthill's vorticity-based 
picture of turbulent boundary layers with the bursting phenomenon and the 
attached-eddy model \citep{gadelhak90}. Using the standard assumptions in the AEM that
%\textcolor{blue}{given outer length scale $h$ and $y^*=y/h,$}
the eddy-intensity function satisfies $I_{xy}(y^*)\sim -Qy^*$ 
for $y^*\ll 1$ and $I_{xy}(y^*)\to 0$ for $y^*\gtrsim 1$\citep{townsend1976structure,woodcock2015statistical},  
\cite{eyink2008} showed that the ensemble-average nonlinear vorticity flux in the AEM for 
turbulent channel and pipe flow with friction velocity $u_\tau$ and outer length $H$ is
\be \langle v\omega_z-w\omega_y\rangle\propto -\frac{Qu_\tau^2}{H}.\ee
The AEM thus predicts the correct outward nonlinear flux of vorticity away from the wall at heights $y>y_p,$ the location of peak Reynolds stress, consistent with some contemporary 
claims that the AEM applies only for that range of wall distances \citep{marusic2019attached}.
The standard AEM, however, certainly does not explain the nonlinear vorticity transport flux toward the wall in the lower part of the log-layer for $y<y_p$ \citep[][\S III.B.2]{eyink2008}. 
The latter phenomenon was the main focus 
of \cite{lighthill1963} and its absence in the standard AEM suggests that a complement is needed. Recently, vorticity-based methods have provided some evidence in support of Lighthill's mechanism, first in a transitional boundary layer by a stochastic Lagrangian analysis \citep{wang2022origin} and next in a fully developed 
turbulent channel flow, both by an Eulerian analysis of vorticity flux \citep{kumar2023flux} and also by Lagrangian analysis \citep{xiang2025origin}.

\iffalse 
The latter verification is more subtle, because it is not completely obvious how 
Lighthill's original Lagrangian argument should be interpreted in terms of Eulerian
vorticity flux. In general agreement with Lighthill's proposal, \cite{kumar2023flux} observed mean vorticity flux away from the wall when conditioning upon positive wall-normal velocity and 
mean vorticity flux toward the wall when conditioning upon negative wall-normal velocity. 
The latter flux direction was termed ``up-gradient'', because it carries more vorticity into 
the near-wall region of concentrated high vorticity, whereas flux away from the wall 
was likewise called ``down-gradient''. In addition, \cite{kumar2023flux} made a 
novel observation of anti-correlation between the vorticity fluxes associated with advection and stretching/tilting of spanwise vorticity, both when conditioned on outflow from the wall 
and also when conditioned on inflow to the wall. 
\fi 

%Although this anti-correlation had not been predicted by \cite{lighthill1963}, a control-volume argument inspired by Lighthill's 
%picture was suggested by \cite{kumar2023flux} to explain this rather puzzling anti-correlation. 

Here we make a definitive advance by calculating conditional averages 
\citep{kim1986structure,adrian1989} of velocity fields and vorticity fluxes 
for positive or negative values of 
wall-normal velocity throughout the entire log-layer. 
Our results for the conditional mean vorticity flux and the vortex lines of the conditional mean fields presented in detail below support the view that Lighthill's mechanism acts in the same manner throughout that entire range, with the conditional structures merely 
growing in scale with increasing distance of the conditioning point from the wall. 
As a preview of these results, we show in figure~\ref{eddy} the mean vortex structures and
vortex lines conditioned at one height $y_c.$ Details of the numerical methods 
are given in Section \ref{sec:num}. 
The results shown in figure~\ref{eddy}(a) are  
conditioned on motion outward from the wall at the point $y_c^+=92.8$ inside the log-layer. The vortex structure has the familiar form of a ``hairpin vortex'', with an 
elevated spanwise head above the conditioning point and streamwise legs near the wall. 
On the other hand, the results shown in figure~\ref{eddy}(b) for 
conditioning on motion inward to the wall at the same point inside the log-layer 
appear quite distinct. The conditional structure appears as a broad 
necklace vortex, or ``shawl vortex'', wrapped around the down-flowing fluid mass.
We remark that these are similar to a smoothed version of the vortex structures visualized in \cite{kumar2023flux}
for a velocity field spectrally filtered to contribute only up-gradient nonlinear vorticity 
flux. Observations qualitatively similar to these will be presented below for conditioning 
points at every wall distance in the log-layer.  
The vortex lines for our two conditions closely 
resemble those for the conditions QD2 and QD4 in the classic work of 
\cite{kim1986structure}, except that they studied momentum transport 
whereas we focus on vorticity transport. 
For a more quantitative discussion, therefore, we must first recall the definition 
of Eulerian vorticity flux which is the basis of our work.

\vspace{-13pt} 
\section{Methods of the Present Study} 

\subsection{Theoretical Methods} \label{sec:prior}

The analysis of the present work relies on the {\it Eulerian vorticity flux tensor} originally 
introduced by \cite{huggins1971dynamical,huggins1994vortex} (see also 
\cite{eyink2008,terrington2021generation,kumar2023flux,kumar2024josephson,du2025ja})
which for incompressible Navier-Stokes is:  
\begin{align} \Sigma_{ij} &= u_i\omega_j-u_j\omega_i 
+\nu\left(\frac{\partial\omega_i}{\partial x_j}-\frac{\partial\omega_j}{\partial x_i}\right).
\label{Sigma} \end{align} 
%\begin{align} h&=p/\rho+|\mathbf{u}|^2/2+Q. \end{align}
The above tensor describes the spatial flux of the $j$th component of vorticity in the $i$th 
coordinate direction, with its anti-symmetry arising from the fact that vortex lines 
cannot terminate in the fluid \citep{terrington2021generation}. Thus, this tensor appears 
as a space-transport term in the local balance equations for the $j$th components of 
vorticity, $\partial_t\omega_j+\partial_i\Sigma_{ij} =0$ for $j=1,2,3.$ The three terms 
in \eqref{Sigma} have transparent physical meaning, with the first representing advective 
transport of vorticity, the second transport by stretching and tilting of vorticity, and the 
third transport by viscous diffusion of vorticity. 

The most important component of the vorticity flux for explanation of drag is $\Sigma_{yz},$ the flux of spanwise $z$-vorticity in the 
wall-normal $y$-direction \citep{kumar2024josephson}. It has been pointed out by many researchers \citep{taylor1932,huggins1994vortex,klewicki2007physical,eyink2008,brown2015vorticity,kumar2023flux}: 
that the mean value of this flux in statistically steady-state 
Poiseuille flow is exactly equal to the pressure-gradient 
in the streamwise $x$-direction, and furthermore is constant in $y$ for pipe and channel flow because 
of stationarity and conservation of vorticity ($u_\tau$ friction velocity, $H$ channel half-width): 
\be \langle \Sigma_{yz}\rangle =\langle v\omega_z- w\omega_y-\nu(\partial_y\omega_z-\partial_z\omega_y)\rangle= \partial_x \langle p\rangle= -u_\tau^2/H. \lb{const-flux} \ee 
As stressed by \cite{huggins1994vortex}, equation \eqref{const-flux} is a classical equivalent of the 
time-average relation of \cite{Josephson65} and \cite{Anderson66} for quantum superfuids, which relates 
drag to vortex motion. Here we always mean $\Sigma_{yz}$ whenever we refer to ``vorticity flux''. 

\begin{figure}
% \centering
\begin{center}
\includegraphics[width=0.8\textwidth]{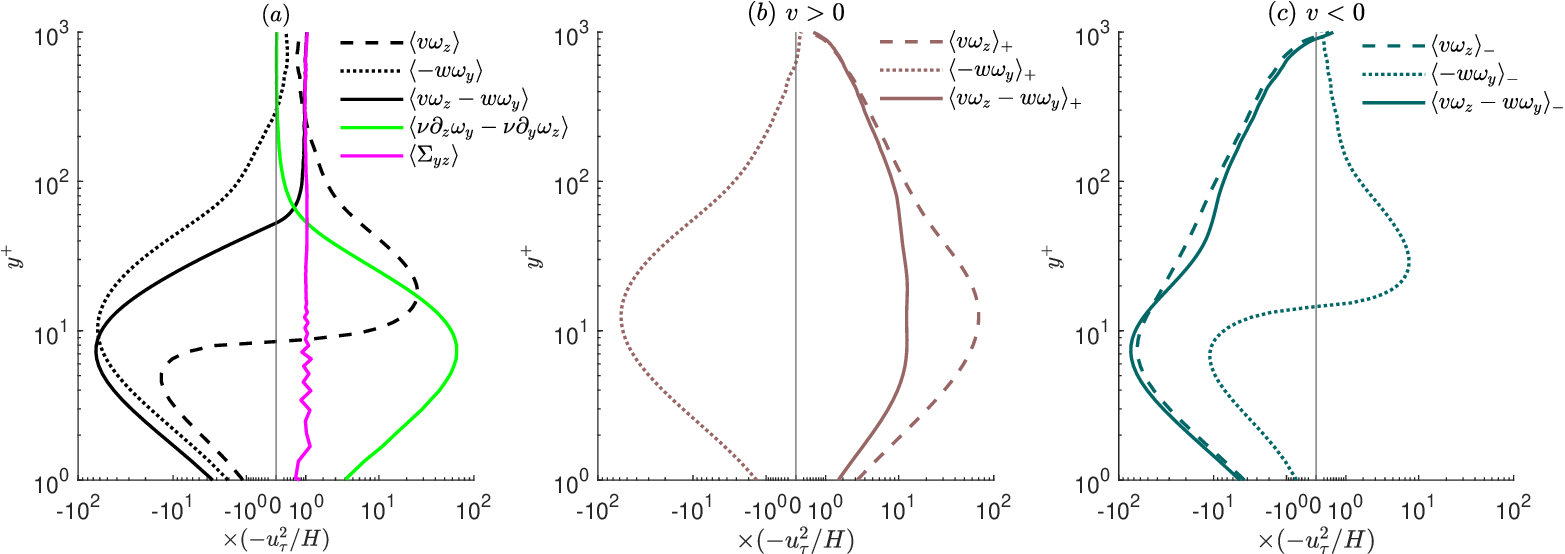}
\end{center}
\caption{Profiles of the mean vorticity flux contributions (a), 
averaged over time and wall parallel planes, plotted as functions of wall distance. The vertical light grey line at zero flux is added to emphasize the signs of the
various contributions.
Conditional averages are plotted in (b) from points where turbulent flow is outward $(v'>0)$ and in (c) where it is inward $(v'<0),$ for the total nonlinear flux and its advective and stretching/tilting parts. The latter two are anti-correlated over the  
log-layer, both for (b) inflow and (c) outflow.}
\label{fig:fluxes} 
\end{figure}

In the prior work by \cite{kumar2023flux}, the constant-flux 
relation \eqref{const-flux} was verified by numerical simulation 
data, together with the observation of \cite{klewicki2007physical,eyink2008,brown2015vorticity} that 
$\langle v\omega_z- w\omega_y\rangle>0$ for $y<y_p$ and $\langle v\omega_z- w\omega_y\rangle<0$ for $y>y_p,$
where $y_p$ is the wall-distance location of the peak Reynolds stress. See Fig.~\ref{fig:fluxes}(a),
which reproduces Fig.~5 in \cite{kumar2023flux} for the $Re_\tau=1000$ channel-flow database where $y_p^+=52.$ Since \cite{lighthill1963} had argued for strong correlation with the wall-normal velocity, \cite{kumar2023flux} calculated also the average fluxes conditioned on $v>0$ and $v<0.$ These conditional means are presented in Fig.~\ref{fig:fluxes}(b),(c),
reorganizing the data from Fig.~7 in \cite{kumar2023flux}. As shown, 
the net nonlinear vorticity flux is ``down-gradient'' or away from the wall for $v>0$ and ``up-gradient'' 
or toward the wall for $v<0.$ As intuitively obvious, the mean advective flux contribution has these same 
signs, but intriguingly Fig.~\ref{fig:fluxes}(b),(c)) show that the mean stretching/tilting contribution has 
the opposite sign throughout the logarithmic layer. \cite{kumar2023flux} 
presented a tentative explanation of this anti-correlation effect based on an assumed geometry of vortex lines
and Lighthill's predicted flow behavior: spanwise converging for $v>0$ and spanwise diverging for 
$v<0.$ The main goal of the present work is to check Lighthill's picture in detail, and also the 
related explanation for anti-correlation, by calculating mean flow 
structures and their vortex lines conditioned on the wall-normal velocity at various points 
throughout the log-layer.

\subsection{Numerical Methods}\label{sec:num} 

We employ direct numerical simulation data of channel flow at 
$Re_{\tau}=1000$ from the Johns Hopkins Turbulence Database (JHTDB) \citep[see][]{jhtdb1,jhtdb_channel}. 
%The right-handed Cartesian coordinate system is used with $x$ streamwise, $y$ wall-normal and $z$ spanwise direction. 
We have used the database cut-out service to download time snapshots of data for the entire channel. Gradients in the spanwise and streamwise directions are then calculated spectrally by FFT, and wall-normal gradients are calculated using seventh-order basis-splines based on the collocation points of the 
original simulation \citep{jhtdb_channel}.
All statistics are thereafter calculated by averaging over wall-parallel planes 
in the $x$ and $z$ directions of homogeneity, as well as over 10 time snapshots. Reflected results from the top half of the channel are included to double the sample size of our averages. 

Our conditional averaging was designed to select points with a local maximum of wall-normal velocity magnitude $|v({\bf x})|$ exceeding some threshold $\alpha v_{rms}.$ We have checked 
that our results do not depend very sensitively upon the choice of $\alpha$ and we present 
results here only for $\alpha=1$.  We may argue for the reasonableness of this choice by 
noting that, for both signs $\pm$ of $v$ and independent of $y$, the set of points with $|v|>v_{rms}$ constitute 10\% of the area of the wall-parallel plane at that $y$-level but contribute about 
$60\%$ of the total vorticity flux for that sign of $v,$ as plotted in Fig.~\ref{fig:fluxes}(b),(c). See 
Appendix,\S \ref{sec:frac}. To make certain that the events in the conditional ensemble are distinct, we set streamwise ($x_w$) and spanwise ($z_w$)
extents of the sampling window for each event. For each sign $\pm$ of $v$ and each $y$-level, 
we then performed sequentially the following steps: (i) identify the point with largest magnitude 
of $|v|$ (and above $v_{rms}$) in the wall-parallel plane, (ii) add the sample of size $x_w\times z_w$ 
centered at that point, (iii) remove from the plane the doubled 2$x_w\times 2z_w$ rectangle centered 
at the point in order to prevent overlap, (iv) find the next point 
with largest magnitude of $|v|$ (and above $v_{rms}$) in the remaining portion 
of the wall-parallel plane, and so forth. The size of the sampling window at each $y$-level
was selected by calculating approximate conditional averages with a linear estimator \citep{adrian1989}
and determining the smallest rectangle to contain the conditional eddy visualized by the 
$\lambda_2$-criterion at a low threshold.

The sizes of the sampling windows and the number of events in the conditional ensembles 
for each sign $\pm$ and for five values of $y=y_c$ distributed through the log-layer 
are given in Table~\ref{tab:num_events}. For decreasing $y_c,$ the sizes of the events 
decrease, as measured by the areas $x_w\times z_w,$ while the total number of events 
increase. In fact, as an {\it a posteriori} justification of our sampling procedure, 
we note that the percentage of the total area occupied both by the outflow ($v_+$) events and by the inflow ($v_-$)
events is about 47\% for each sign, independent of $y_c.$ This $y$-independence of the 
area fraction is expected of the ``representative eddies'' in the attached-eddy 
model, but notice that such independence holds here for the $v_-$ events or ``sweeps'' as well
as for the $v_+$-events or ``ejections". In fact, the events in our two conditional 
ensembles cover together nearly the entire area of the wall-parallel plane at each 
$y$. Note that for all $y$-levels except the largest there are slightly more $v_+$
events than $v_-$ events at the same threshold, reflecting the well-known 
asymmetry in strengths of ``ejections'' versus ``sweeps''
\citep{willmarth_lu_1972,kim1986structure,hutchins2011three,lozano2012three}.

\begin{figure}
\begin{center}
{\includegraphics[width=0.6\textwidth]{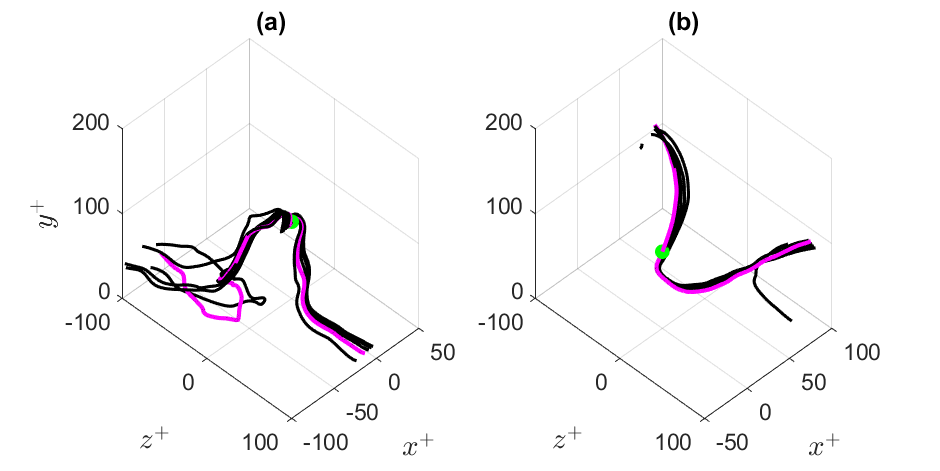}}
\end{center}
\caption{ Instantaneous vortex lines for (a) outflow event (b) inflow event, both in the 
vicinity of a local maximum
of the wall-normal velocity at $y^+=92.8$. The unique vortex line passing through the conditioning point is marked in magenta.}
\label{fig:instant} 
\end{figure}

To provide some intuition about the events selected by our sampling procedure, 
we plot in Figure~\ref{fig:instant} one event in the  $v_+$ ensemble and another 
in the $v_-$ ensemble. To visualize these events, we have followed 
\cite{kim1986structure} in drawing the unique vortex line passing through
the conditioning point and also nearby vortex lines. 
The resulting bundle of vortex lines for the outflow event in  Figure~\ref{fig:instant}(a)
is easily recognizable as a ``hairpin vortex'', while the bundle for the inflow event in  Figure~\ref{fig:instant}(b) is instead an ``inverted hairpin''. These two events are 
for the same wall-distance $y^+_c=92.8$ as the mean structures plotted 
in Fig.~\ref{eddy}, where the means are obtained by averaging over 
the entire conditional ensembles, hereafter denoted as $\langle\cdot\rangle_{+,y_c}$ 
and $\langle\cdot\rangle_{-,y_c}$, respectively. (We omit $y_c,$ if it is clear in context). 
For more such events, see Appendix \S \ref{sec:lines}. The two events plotted 
in Fig.~\ref{fig:instant} have the largest magnitudes of $|v({\bf x})|$ for the given time, sign, 
and $y$-level. 

\iffalse 
\textcolor{blue}{Conditional averages are calculated by first identifying a grid point at each wall normal location satisfying two conditions: 1) the point should be a local maximum (corresponding to outflow from the wall, called $v_+$) / minimum (corresponding to inflow towards the wall, called $v_-$) of wall normal velocity; 2) the magnitude of the velocity should be greater than than $v_{rms}$ at that wall-normal location. Instantaneous vortex lines associated with both an outflow and an inflow event are illustrated in Fig.~\ref{fig:instant}, the conditioning point being marked with a green point.
The number of inflow and outflow events sampled at each wall normal location is listed in Table \ref{tab:num_events}. Velocity and flux fields from a sub-region (or averaging window) of the channel centered at the point are added to calculate the conditional mean. Care is taken to ensure that the averaging windows associated with different points do not overlap. The streamwise $(x_w)$ and spanwise $(z_w)$ extents of the averaging window are listed in Table \ref{tab:num_events}. In the wall-normal direction, the averaging window extends from the wall to the channel center. }  \fi

\begin{table}
  \centering
  \begin{tabular}{lccc}
    \textbf{$y_c^+$} & \textbf{No. of $v_{+}$ events } & \textbf{No. of $v_{-}$ events} & \textbf{$x_w^+ \times z_w^+$}\\
    ~~39 & 20{,}593 & 20{,}366 &  $515 \times208$ \\
    ~~52 & 16{,}954 & 16{,}775 & $515 \times 257$ \\
    ~~93 & 12{,}143 & 12{,}033 & $614 \times 307$\\
    197 & ~~7{,}925 & ~~7{,}849 & $712 \times 405$\\
    298 & ~~4{,}660 & ~~4{,}720 & $810 \times 601$\\
  \end{tabular}
  \caption{Number of outflow and inflow events sampled at various wall normal locations along with the streamwise 
  ($x_w$) and spanwise ($z_w$) extent of the sampling window.}
  \label{tab:num_events}
\end{table}

\vspace{-10pt} 
\section{Results of Conditional Averaging}\label{sec:results} 

We present our results first for averages conditioned on outflow (\S\ref{sec:out}) and then 
on inflow (\S\ref{sec:in}). The log-layer in the $Re_\tau=1000$ database extends over the range 
$30\lesssim y^+\lesssim 300$ and we present results in the first two sections for a single height 
$y_c^+=92.8,$ roughly at the geometric mean of the log-layer. Finally, we consider (\S\ref{sec:scale})
the variation of our numerical results with wall-distance $y^+$ and the evidence for a scale-hierarchy.

\subsection{Outflow from the Wall}\label{sec:out}

\begin{figure}
\begin{center}
{\includegraphics[width=\textwidth]{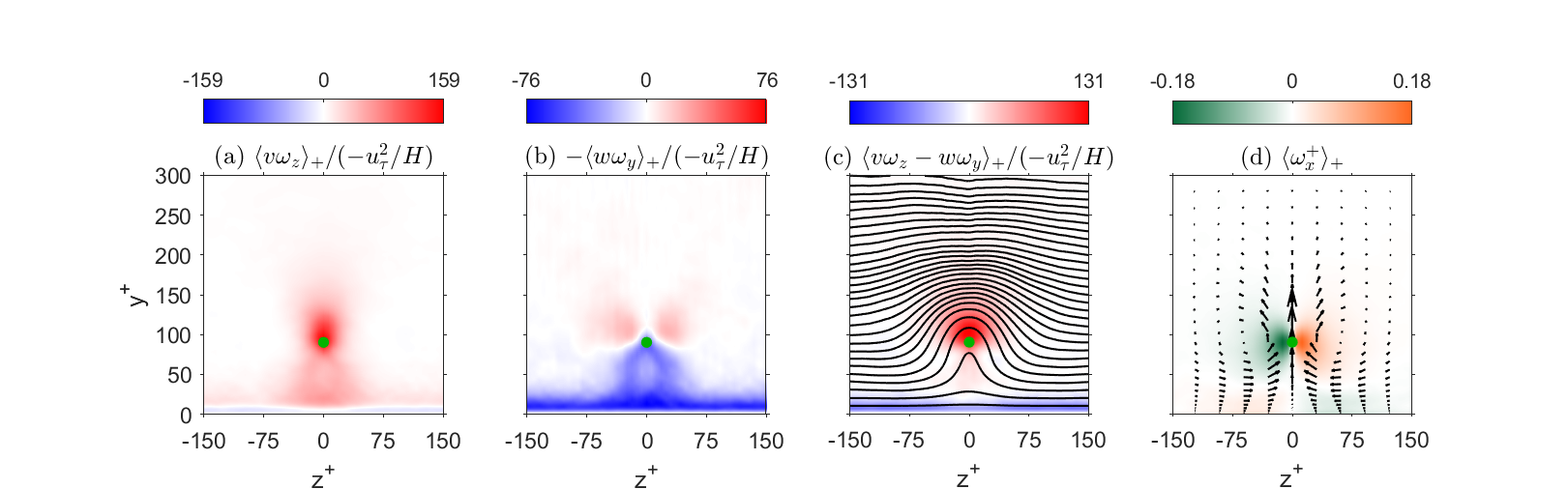}}
\end{center}
\caption{Conditional mean fields {in the plane of the conditioning point} for the outflow event at $y_c^+=92.8$, colored by (a) flux due to the convective term, (b) flux from the stretching/tilting term, (c) total nonlinear flux, {(d)} streamwise vorticity. \iffalse Colorbars are asymmetric in (a)-(c) to emphasize sign changes. \fi Also depicted are (c) vortex lines and (d) quivers showing in-plane velocity. A green dot marks the conditioning point.}
\label{outflow} 
\end{figure}

In Figure~\ref{outflow} we plot the contributions to the mean vorticity flux 
and the mean flow for the conditional average  $\langle \cdot\rangle_{+,y_c}$
at $y_c^+=92.8.$ The first panel 
Fig.~\ref{outflow}(a) plots the mean advective flux under this condition, which is straightforwardly 
down-gradient throughout most of the domain. There is only a very narrow layer at $y^+\lesssim 10$
where the convective flux contribution is up-gradient. The latter effect was explained in 
\cite{kumar2023flux} by the correlation between weakened spanwise vorticity and upward motion in this 
near-wall region. On the other hand, the second panel, Fig.~\ref{outflow}(b), which plots the 
mean stretching/tilting contribution shows strong up-gradient flux below the conditioning 
point, especially near the wall, and a weaker down-gradient flux in the region above the 
conditioning point. The main goal of this subsection is to explain these observations.

The most significant clues to the correct explanation are in the remaining panels. 
Fig.~\ref{outflow}(c) plots in the $yz$-plane of the conditioning point the vortex lines 
of the mean flow. These show the ``hairpin'' structure of vortex lines already evidenced by the 
3D plot in Fig.~\ref{eddy}(a). Furthermore, Fig.~\ref{outflow}(d) shows that the conditional 
flow corresponds to a large-scale ``ejection'' between a pair of counter-rotating streamwise vortices. 
Below these vortices are oppositely-oriented streamwise rollers, 
illustrating Lighthill's remark in the opening quotation of this paper about ``generating new vorticity at the surface'' due to the stick b.c. at the wall. See Figure II.19 in \cite{lighthill1963}.

These characteristics were exactly those predicted by \cite{lighthill1963} to explain the up-gradient 
flux due to weakening of vortex lines and 
{were} also the ingredients of the control-volume argument 
by \cite{kumar2023flux} to explain the anti-correlation between advection and stretching contributions
during outflow. We repeat this argument in Fig.\ref{outflow_sketch}(a), with attention on the 
bottom line, taken as representative of all lines below the conditioning point. Because of the spanwise 
converging flow, the product $-w\omega_y>0$ gives an up-gradient transport into the grey-shaded
control-volume, representing the loss of spanwise vorticity of the lifted vortex line. 
This is precisely Lighthill's mechanism, based on converging flow that compresses and weakens 
the upward-moving vortex lines. Note from Fig.~\ref{outflow}(d) that the mean flow is converging 
all the way up to the conditioning point. Thus, not only does the conditional mean flow support 
Lighthill's mechanism near the wall, but it also shows that Lighthill's mechanism can explain the 
observed anti-correlation at the conditioning point and in a region below it.%everywhere below the conditioning point.

Note, however, that the flow above the conditioning point is instead diverging, 
because of the counterflow required by incompressibility but neglected in the considerations
of \cite{lighthill1963}. His reasoning would suggest that rising vortex lines in this 
region are being stretched and strengthened. This effect is associated to the product $-w\omega_y<0$ 
for the upper vortex line above the conditioning point in Fig.\ref{outflow_sketch}(b),
which corresponds to down-gradient flux out of the top surface of the grey-shaded control volume
and into the lifted vortex head. In fact, down-gradient flux from stretching/tilting is indeed 
observed in this region in Fig.~\ref{outflow}(b). \iffalse , although far above the conditioning point and quite weak.
The conditional eddy overestimates the effect, as it approximates the conditional stretching 
$\langle\omega_y w\rangle_+$ by $\langle\omega_y\rangle_+\langle w\rangle_+,$ neglecting correlations. 
See Appendix, Fig.~\ref{fluxcond}b, which plots the latter. The quantities $\omega_y$ and $w$ seem to have 
magnitudes weakly anti-correlated just above the conditioning point, but strongly correlated below. 
\fi 

Finally, together with the vortex lines in Fig.~\ref{outflow}(c) we have plotted the conditional mean 
nonlinear vorticity flux from both advection and stretching/tilting. This total nonlinear flux 
is down-gradient everywhere 
except near the wall, because the advection term is generally stronger than the stretching term. However, Lighthill's mechanism dominates near the wall,
producing net up-gradient vorticity flux toward the wall.

{\begin{figure}
\centering
\includegraphics[width=0.8\textwidth]{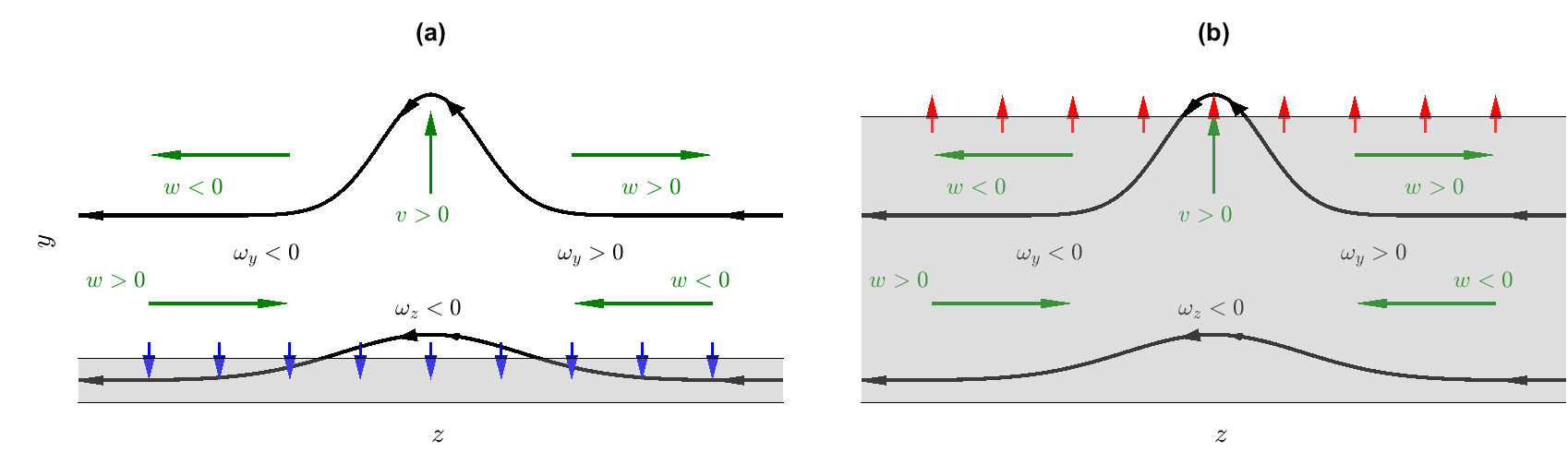}
\caption{Control volume analysis of outflow away from the wall illustrating the stretching contribution to spanwise vorticity balance for (a) Lighthill region, (b) counter-flow region. Black lines with arrows 
represent vortex lines and green arrows mark the directions of local velocity components. Blue arrows 
in (a) at the boundary of the relevant control volume, shaded grey, represent up-gradient flux from the stretching-tilting term into the volume, while red arrows in (b) represent down-gradient flux out of the volume.}
\label{outflow_sketch}
\end{figure}}

\subsection{Inflow to the Wall}\label{sec:in}

\begin{figure}
%\centering
\includegraphics[width=\textwidth]{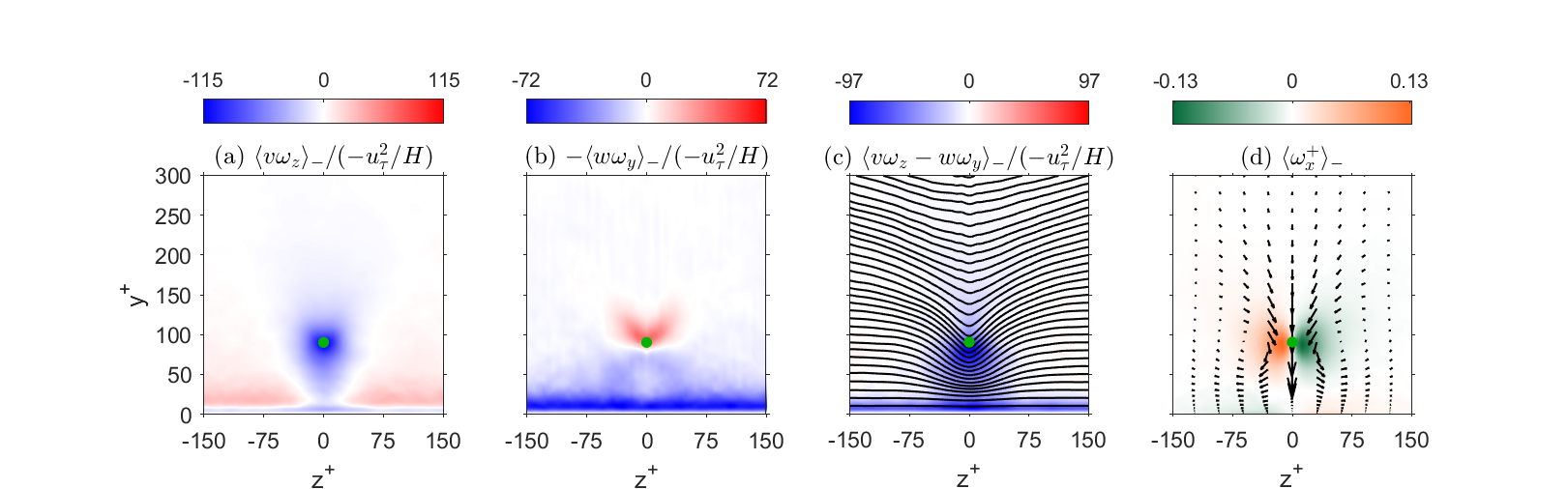}
\caption{Conditional mean fields in the plane of the conditioning point for the inflow event at $y_c^+=92.8$, colored by (a) flux due to the convective term, (b) flux from the stretching/tilting term, (c) total nonlinear flux, (c) streamwise vorticity. Also depicted are (c) vortex lines and (d) quivers showing in-plane velocity. 
A green dot marks the conditioning point.}
\label{inflow} 
\end{figure}

In Figure~\ref{inflow} we plot the analogous contributions to the mean vorticity flux 
and the mean flow for the conditional average $\langle\cdot\rangle_{-,y_c}$ 
at $y^+_c=92.8.$ The advective flux 
plotted in Fig.~\ref{inflow}(a) is again straightforwardly up-gradient near the conditioning point
where the mean flow is toward the wall, but down-gradient below and to the sides. The latter sign
can be explained as a counter-flow effect: see the flow vectors in Fig.~\ref{inflow}(d) directed away
from the wall in this region. The up-gradient advective flux in the layer 
$y^+\lesssim 10$ was already observed in \cite{kumar2023flux} and explained there by 
the correlation between strengthened spanwise vorticity and downward motion in the very near-wall region. 
The second panel, Fig.~\ref{inflow}(b), plots the  mean stretching/tilting contribution and shows 
down-gradient flux in a small region at and just above the conditioning point.
%and also far above. 
On the other hand, close to the wall the stretching contribution is strongly up-gradient. In this 
subsection we develop an explanation of these various observations. 

As before, the most significant pieces of information are in the remaining panels of 
Fig.~\ref{inflow}. The vortex lines of the mean flow plotted in Fig.~\ref{inflow}(c) 
over the $yz$-plane of the conditioning point and plotted also for 3D in Fig.~\ref{eddy}(b)
have the form of ``inverted hairpins''.  Furthermore, Fig.~\ref{inflow}(d) shows that 
the conditional flow corresponds to a large-scale ``sweep'' between a pair of counter-rotating 
streamwise vortices, opposite in orientation to the pair in Fig.~\ref{outflow}(d), and again 
with induced streamwise rollers of the opposite sign near the wall. 
These characteristics are precisely those predicted by \cite{lighthill1963} 
for in-flow to the wall, 
with a spanwise diverging flow beneath the conditioning point. To understand the up-gradient 
transport of the stretching term in that region, we can use a control-volume analysis which 
assumes a field-line geometry like that illustrated in Figs.~\ref{eddy}(b), \ref{inflow}(c). See the 
bottom line sketched in Fig.~\ref{inflow_sketch}(a). Because of the spanwise diverging 
flow, the product $-w\omega_y>0$ gives an up-gradient transport into the grey-shaded
control-volume, representing the gain of spanwise vorticity of the vortex line. 
This is exactly Lighthill's mechanism, based on diverging flow that stretches and 
strengthens the downward-moving vortex line.

{\begin{figure}
\centering
\includegraphics[width=0.8\textwidth]{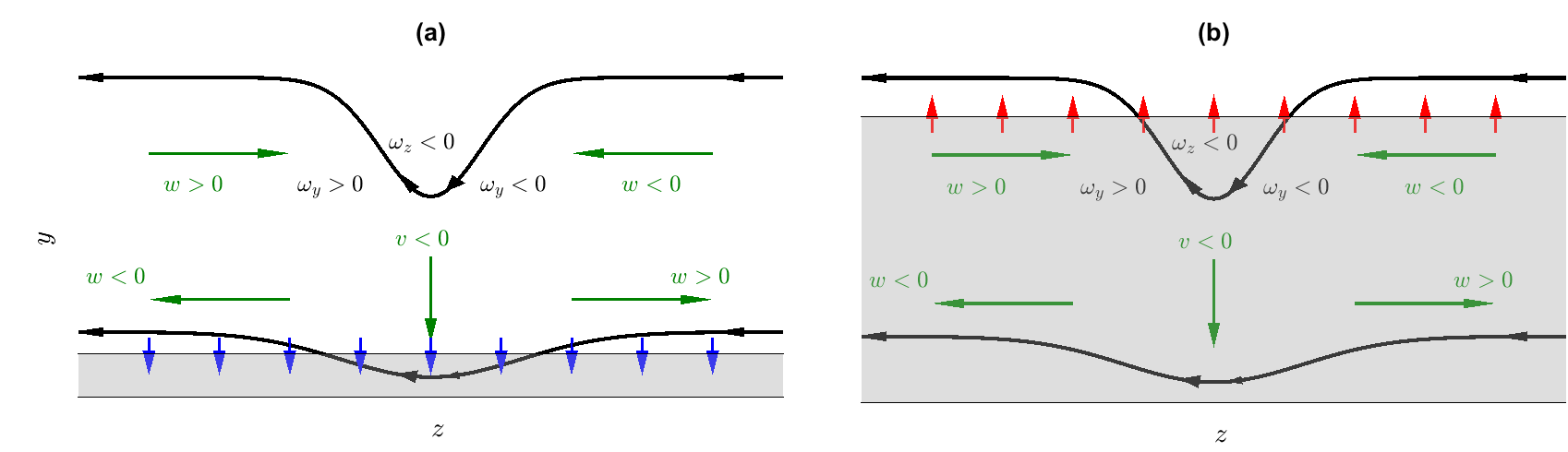}
\caption{ Control volume analysis of inflow towards the wall illustrating the stretching contribution to spanwise vorticity balance for (a) Lighthill region, (b) counter-flow region. 
Conventions for lines, arrows and their colors are the same as in Fig.~\ref{outflow_sketch}.
}
\label{inflow_sketch}
\end{figure}}

As with the out-flow case, however, a recirculation appears that was not considered by Lighthill 
and that leads now to a converging flow above the conditioning point. 
See Fig.~\ref{inflow}(d). Lighthill's reasoning would suggest here that the down-moving 
vortex lines in this region are compressed and weakened. This effect is associated to the 
product $-w\omega_y<0$  for the upper vortex line above the conditioning point in 
Fig.\ref{inflow_sketch}(b), which corresponds to down-gradient flux out of the top surface 
of the grey-shaded control volume and weakening of the vortex line. This argument
thus suggests that the anti-correlation between advective and stretching contributions 
is in fact due to the counterflow in the case that $v'<0$ and the control-volume picture in Fig.\ref{inflow_sketch}(b)
corrects that of \cite{kumar2023flux}, which erroneously posited an upward-bent hairpin-type line-geometry.  
\iffalse 
Note that the down-gradient flux from the stretching term appears only very near the 
conditioning point and also far above it, but (at least at this $y$-value) a sizable region 
of very weak up-gradient flux appears in the region of converging counterflow moderately 
above the conditioning point. Thus, the previous explanation based on the conditional flow 
overestimates the down-gradient flux above the conditioning point because it approximates $\langle\omega_y w\rangle_-$ 
by $\langle\omega_y\rangle_-\langle w\rangle_-,$ neglecting correlations. See Appendix, Fig.~\ref{fluxcond}d. Just as for outflow, 
the magnitudes of $\omega_y$ and $w$ appear weakly anti-correlated somewhat above the conditioning 
point but strongly correlated below it.  \fi 

We plot also  in Fig.~\ref{inflow}(c) the conditional mean nonlinear vorticity flux from 
both advection and stretching/tilting, together with the vortex lines. This total nonlinear 
flux is up-gradient everywhere,  
%near the conditioning point and down-gradient throughout the rest of the flow away from the wall, 
because the advection term is stronger than the stretching term near the conditioning point. 
However, Lighthill's mechanism again dominates 
near the wall, so that the net flux is likewise up-gradient close to the wall. 
As already emphasized by \cite{lighthill1963}, both outflow and inflow act to concentrate vorticity 
near the wall, the first through weakening and the second through strengthening of the advected vorticity.

Our explanation of the conditionally averaged vorticity fluxes 
$\langle v\omega_z\rangle_\pm,$ $-\langle w\omega_y\rangle_\pm$ based 
on the conditionally averaged fields is {\it a priori} valid only for the 
fluxes $\langle v\rangle_\pm\langle\omega_z\rangle_\pm,$ $-\langle w\rangle_\pm\langle\omega_y\rangle_\pm$ of the conditional eddies themselves. 
The success of this explanation requires that correlations 
of the two fluctuating factors must be rather small in the conditional ensembles. 
We have directly verified the small size of the Pearson correlation 
coefficients (see Appendix, \S \ref{sec:corr}), but a complete physical justification remains open. A possible  
explanation is that the velocities $v,$ $w$ are mainly large-scale quantities
while the vorticities $\omega_z,$ $\omega_y$ are mainly small-scale quantities 
and fluctuations of the two sets of variables are thus naturally uncorrelated 
due to scale-separation \citep[][Section 8.2]{tennekes1972first}. {However, statistical correlations
between these variables are obviously required for turbulent nonlinear transport of vorticity. One may infer 
from the small Pearson coefficients observed in the conditional ensembles that these correlations 
are mainly linked to the direction of the wall-normal velocity, as intuited by Lighthill. This fact provides strong 
{\it a posteriori} justification for our procedure of conditioning on the wall-normal velocity.}

\begin{figure*}
\centering
{\includegraphics[width=\textwidth]{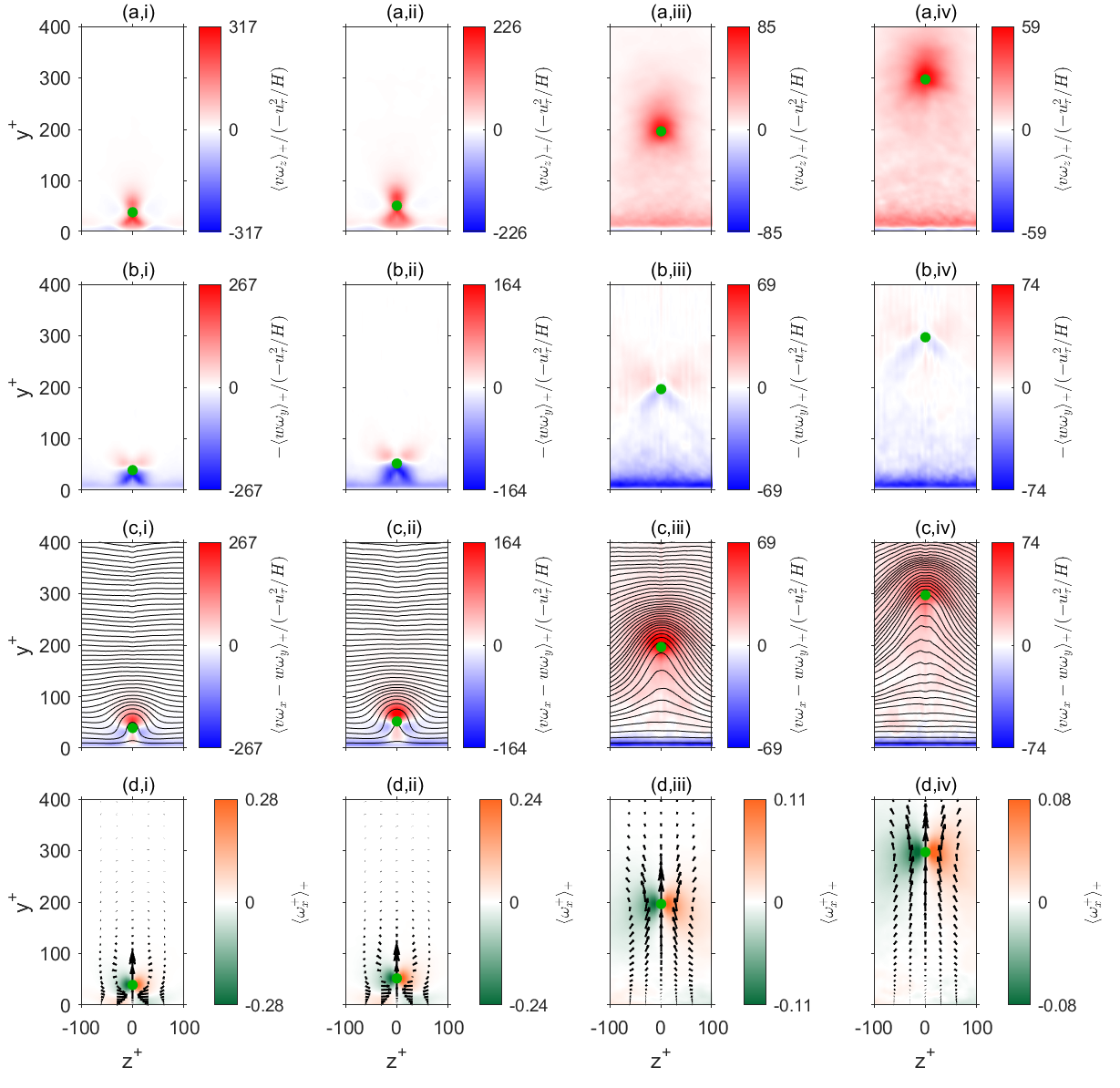}}
\caption{{Conditional fields for outflow events at (a-d,i) $y_c^+=39,$ (a-d,ii) $y_c^+=52,$ (a-d,iii) $y_c^+=197,$ 
(a-d,iv) $y_c^+=298,$ colored by (a,i-iv) flux due to the convective term, (b,i-iv) flux from the stretching/tilting term, 
(c,i-iv) nonlinear flux and  vortex lines in black, (d,i-iv) streamwise vorticity and in-plane velocity as quivers. 
Green dots mark conditioning points.}}
\label{outflow_compiled} 
\end{figure*}

\begin{figure*}
\centering
{\includegraphics[width=\textwidth]{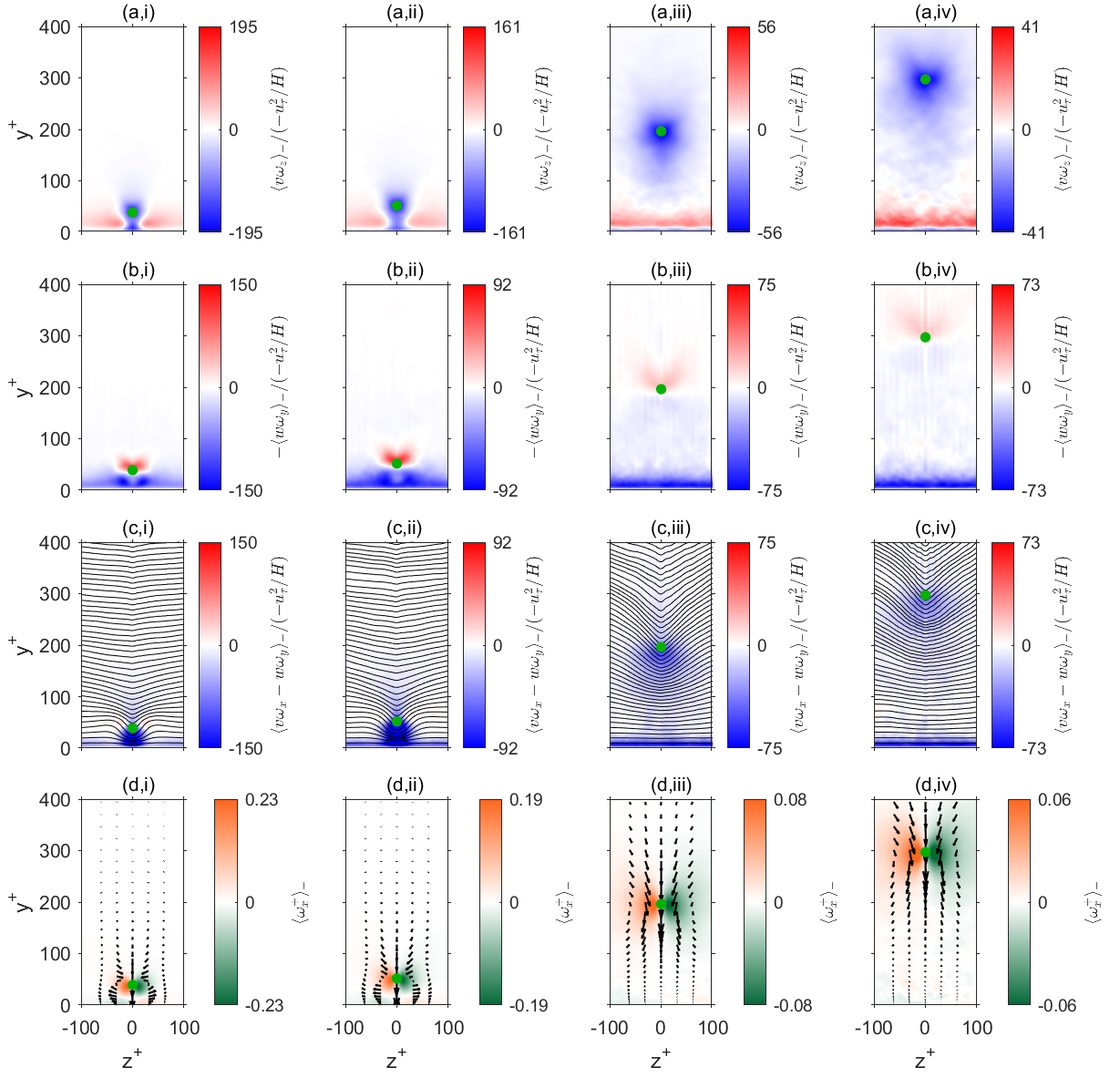}}
\caption{{ Conditional fields for inflow events at (a-d,i) $y_c^+=39,$ (a-d,ii) $y_c^+=52,$ (a-d,iii) $y_c^+=197,$ 
(a-d,iv) $y_c^+=298,$ colored by (a,i-iv) flux due to the convective term, (b,i-iv) flux from the stretching/tilting term, 
(c,i-iv) nonlinear flux and  vortex lines in black, (d,i-iv) streamwise vorticity and in-plane velocity as quivers. 
Green dots mark conditioning points. }}
\label{inflow_compiled} 
\end{figure*}

%\vspace{-10pt}
\subsection{Scaling with Wall Distance}\label{sec:scale}

The results presented previously for conditioning point $y^+_c=92.8$ hold for all 
points within the log-layer.
%, which for our $Re_\tau=1000$ database comprises the range $30\lesssim y^+\lesssim 300.$ 
\iffalse
To show this, we plot in Figure \ref{loglayer}  
the conditional mean vorticity flux and vortex lines, for four $y_c$-values 
in the log-layer, for both outflow and inflow, analogous to the 
plots in Fig.~\ref{outflow}(c) and in Fig.~\ref{inflow}(c), respectively, for  $y^+_c=92.8.$
\fi 
{We show  in Figure \ref{outflow_compiled} at four $y_c$-values in the log-layer results for outflow
analogous to those in Fig.~\ref{outflow} for  $y^+_c=92.8,$ and likewise in Figure 
\ref{inflow_compiled} at the same four $y_c$-values results for inflow analogous to those in Fig.~\ref{inflow}.
The essential features are the same for all $y_c$ as for $y^+_c=92.8,$ with each panel of 
Fig.~\ref{outflow_compiled} showing an ejection between a pair of counter-rotating streamwise vortices
and each panel of Fig.~\ref{inflow_compiled} a sweep between a counter-rotating vortex pair of the opposite 
orientation.} \iffalse 
Note for all $y_c$-values the region of strong up-gradient 
nonlinear vorticity transport near the wall, which can again be explained by Lighthill's mechanism. \fi 
The primary change with increasing $y_c$ is the increased scale of the conditional 
mean events, along with decreasing magnitude of the mean fluxes. 
\iffalse 
The same increasing scale is apparent in the plots analogous to the other panels in Fig.~\ref{outflow} and in Fig.~\ref{inflow} but now for $y_c$ throughout
the log-layer: see Appendix, \S \ref{sec:heights}. 
\fi 
{Only minor qualitative changes with $y_c$ appear, such as a small region of down-gradient transport 
near the wall in the total vorticity flux in Fig.~\ref{outflow_compiled} for outflows at $y_c^+\lesssim 52$
and a single connected region of down-gradient transport near the wall for the advective flux 
in Fig.~\ref{inflow_compiled} for 
inflows at $y_c^+\gtrsim 197$. Except for these small differences, the plots are nearly the same for 
all values of $y_c.$} 
These results support the conjecture of \cite{lighthill1963} that vorticity transport through the log-layer is a ``cascade process'' sustained 
by a scale-hierarchy of vortex structures.  

Additional evidence for scale similarity is provided in Figure \ref{area}, which shows conditional mean eddies for outflow and for inflow at various conditioning points
with heights $y_c$ selected throughout the log-layer, analogous to the structures 
plotted in Fig.~\ref{eddy} for $y^+_c=92.8.$ These eddies are defined by the $\lambda_2$-criterion for the conditional {mean} fields with a threshold {$\lambda_2=-6 u_\tau ^2 /y^2_c$} that is scaled with $u_\tau$ and $y_c.$
%and furthermore are plotted in Fig.~\ref{area} with all lengths rescaled by $y_c.$ 
Note that the structures are not strongly sensitive to the prefactor
$6$ in the $\lambda_2$ threshold and that 3D versions of all images are available via JFM Notebooks through links provided in the figure caption. 
The plots suggest that the conditional eddies change with increasing
$y_c$ chiefly through their streamwise extents decreasing with respect to their spanwise and wall-normal extents, while the latter scale linearly with $y_c.$ These 
observations can be quantified by calculating the side-lengths $X,Y,Z$ for bounding 
boxes of the conditional eddies, which are plotted in Fig.~\ref{boxsize} 
rescaled by $y_c.$ {As observed for both outflow ($v_+$) and inflow ($v_-$), the 
quantities $Y^\pm/y_c$ and $Z^\pm/y_c$ are nearly constant while streamwise size $X^\pm/y_c$ shows some decrease in value. Note also that the extents of outflow eddies are larger than inflow eddies, in part due to the stronger intensity of outflow events. }%are nearly constant $\simeq 1.5 $ for $y_c$ increasing through the log-layer, while $X^\pm/y_c$ slowly decrease from $\simeq 5$ to the same common value $\simeq 1.5.$

{\begin{figure}
\centering
\includegraphics[width=\textwidth]{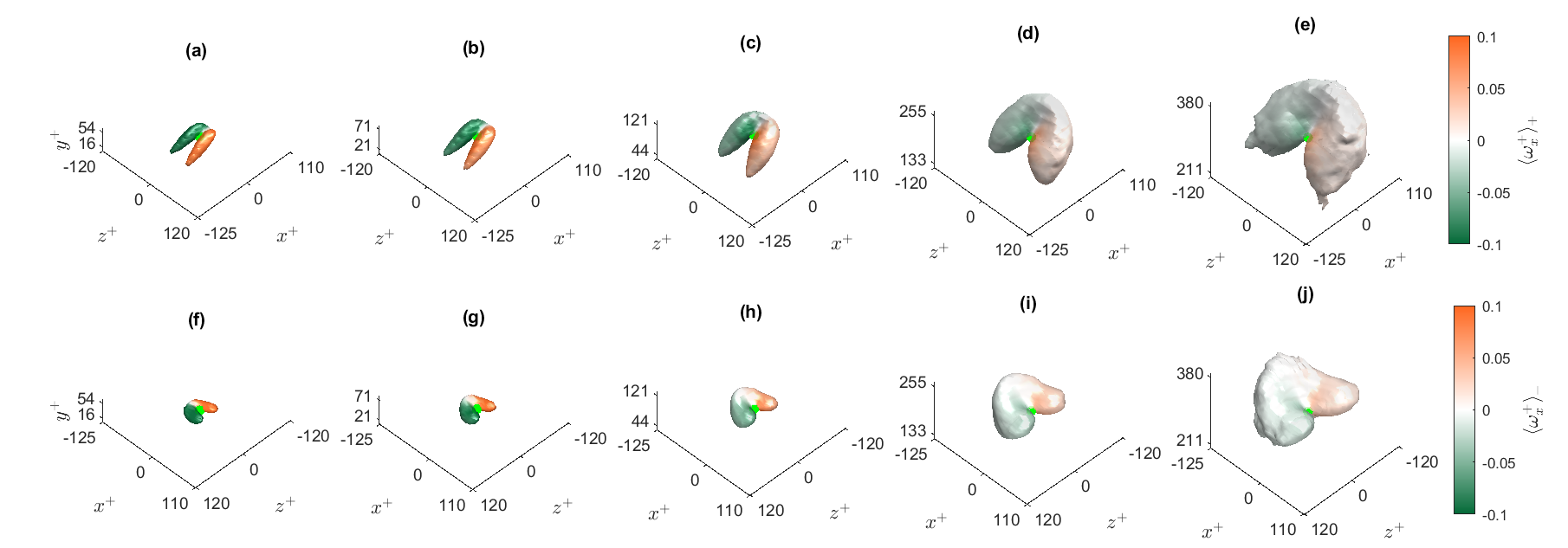}
\caption{{
Outflow eddies (top), inflow eddies (bottom) are illustrated for the conditioning point at (a,f) $y_c^+=39$, (b,g) $y_c^+=52$, (c,h) $y_c^+=92.8$, (d,i) $y_c^+=197$, (e,j) $y^+=298$. The isosurfaces are shown at $\lambda_2 =-6u_\tau ^2 /y_c^2 $. 3D versions of the eddies sketched in this figure, as well as corresponding streamlines for the outflow event are available to view by clicking on \href{https://cocalc.com/share/public_paths/cdee9e3cf9ce3d6804f9ed19f858a04b230f4250/fig10/eddy_streamline_outflow_yplus_039.html}{(a)}, 
\href{https://cocalc.com/share/public_paths/cdee9e3cf9ce3d6804f9ed19f858a04b230f4250/fig10/eddy_streamline_outflow_yplus_053.html}{(b)},
\href{https://cocalc.com/share/public_paths/cdee9e3cf9ce3d6804f9ed19f858a04b230f4250/fig10/eddy_streamline_outflow_yplus_093.html}{(c)}, 
\href{https://cocalc.com/share/public_paths/cdee9e3cf9ce3d6804f9ed19f858a04b230f4250/fig10/eddy_streamline_outflow_yplus_197.html}{(d)},
\href{https://cocalc.com/share/public_paths/cdee9e3cf9ce3d6804f9ed19f858a04b230f4250/fig10/eddy_streamline_outflow_yplus_298.html}{(e)},
and for the inflow event at 
\href{https://cocalc.com/share/public_paths/cdee9e3cf9ce3d6804f9ed19f858a04b230f4250/fig10/eddy_streamline_inflow_yplus_039.html}{(f)},
\href{https://cocalc.com/share/public_paths/cdee9e3cf9ce3d6804f9ed19f858a04b230f4250/fig10/eddy_streamline_inflow_yplus_053.html}{(g)},
\href{https://cocalc.com/share/public_paths/cdee9e3cf9ce3d6804f9ed19f858a04b230f4250/fig10/eddy_streamline_inflow_yplus_093.html}{(h)},
\href{https://cocalc.com/share/public_paths/cdee9e3cf9ce3d6804f9ed19f858a04b230f4250/fig10/eddy_streamline_inflow_yplus_197.html}{(i)},
\href{https://cocalc.com/share/public_paths/cdee9e3cf9ce3d6804f9ed19f858a04b230f4250/fig10/eddy_streamline_inflow_yplus_298.html}{(j)}.
 The code to generate outflow eddies is available \href{https://cocalc.com/share/public_paths/cdee9e3cf9ce3d6804f9ed19f858a04b230f4250/fig10/plot_outflow_eddies_streamlines.ipynb}{here} and to generate inflow eddies is available \href{https://cocalc.com/share/public_paths/cdee9e3cf9ce3d6804f9ed19f858a04b230f4250/fig10/plot_inflow_eddies_streamlines.ipynb}{here}.}
}
\label{area}
\end{figure}}

 {\begin{figure}
\centering
\includegraphics[width=0.8\textwidth]{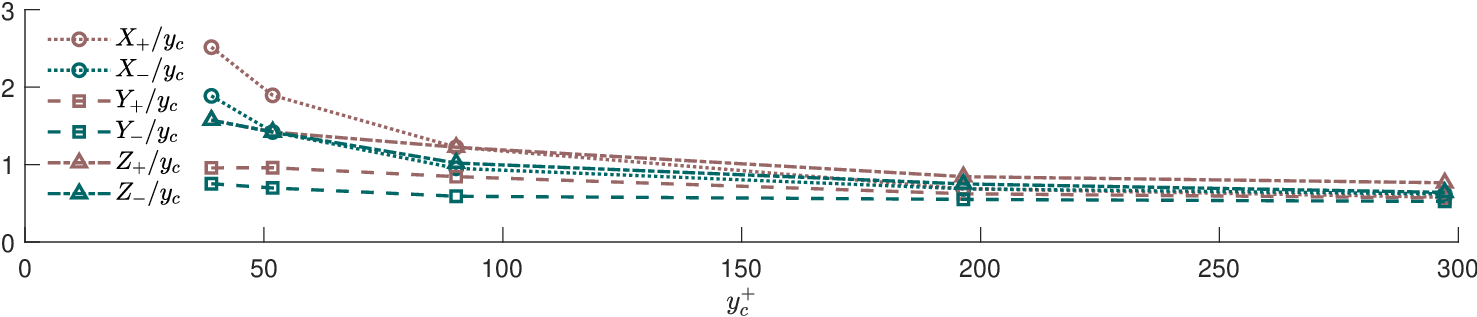}
\caption{Streamwise $(X_+,X_-)$, wall-normal $(Y_+,Y_-)$ and spanwise $(Z_+,Z_-)$ sizes of outflow and inflow conditional eddies scaled by the wall normal location of the conditioning point. 
}
\label{boxsize}
\end{figure}}

{\begin{figure}
\centering
\includegraphics[width=0.8\textwidth]{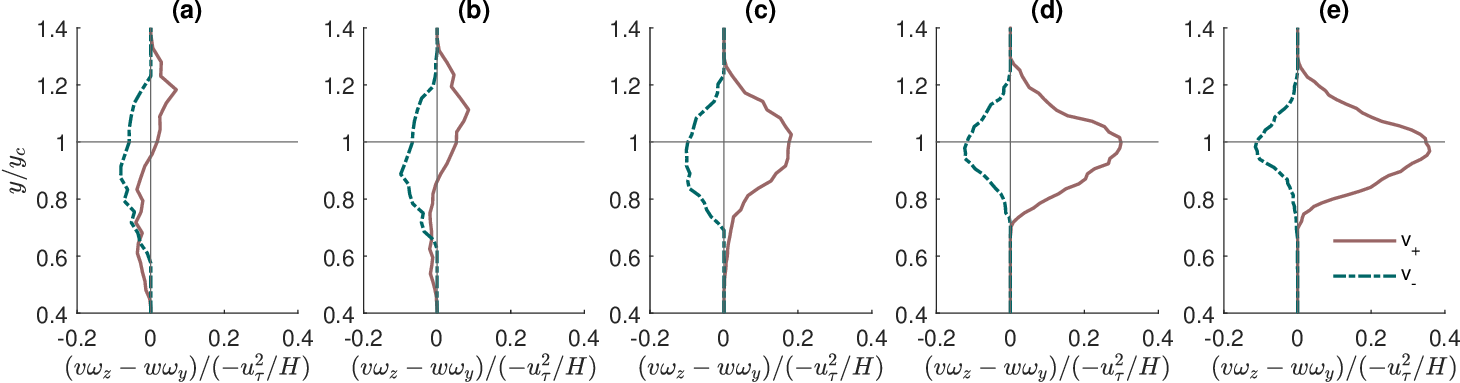}
\caption{{Flux contributions from conditional outflow and inflow eddies, for the conditioning point at (a) $y_c^+=39$, (b) $y_c^+=52$, (c) $y_c^+=92.8$, (d) $y_c^+=197$, (e) $y^+=298$. 
}}
\label{num}
\end{figure}}

{\begin{figure}
\centering
\includegraphics[width=\textwidth]{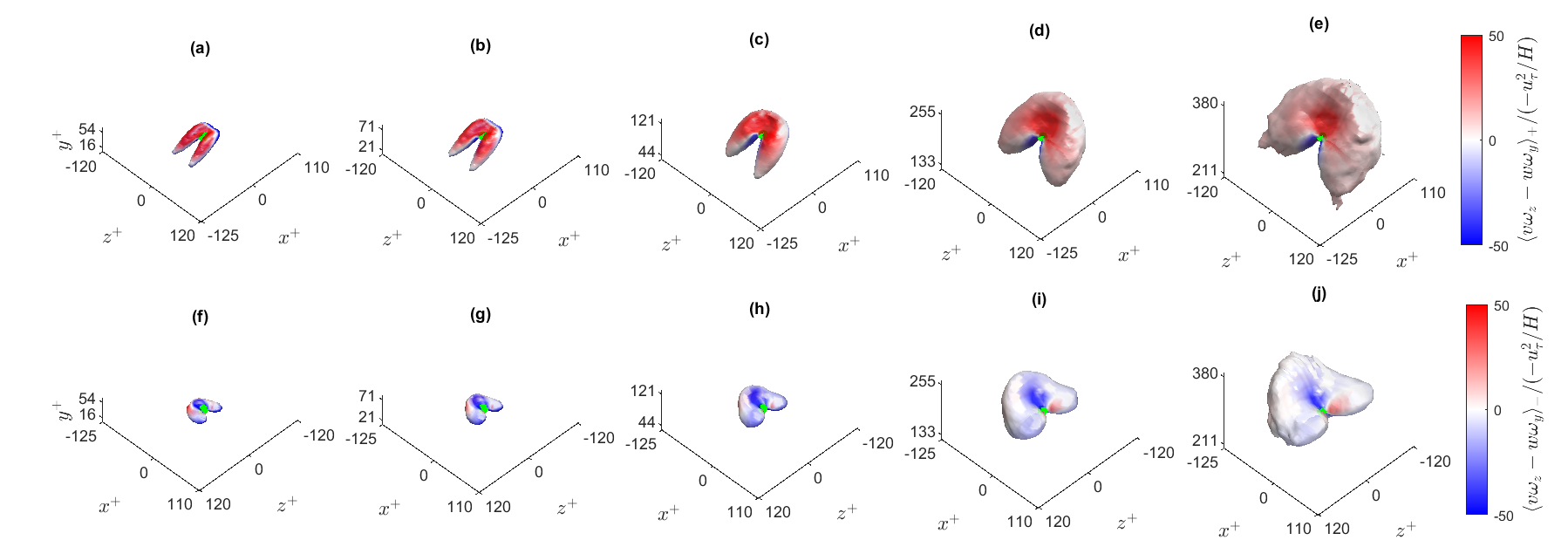}
\caption{ {Outflow eddies (top), inflow eddies (bottom) are illustrated for the conditioning point at (a,f) $y_c^+=39$, (b,g) $y_c^+=52$, (c,h) $y_c^+=92.8$, (d,i) $y_c^+=197$, (e,j) $y^+=298$. The isosurfaces are shown at $\lambda_2 =-6u_\tau ^2 /y_c^2 $ and are colored by the associated nonlinear vorticity flux. 3D versions of the eddies sketched in this figure, as well as corresponding streamlines for the outflow event are available to view by clicking on \href{https://cocalc.com/share/public_paths/cdee9e3cf9ce3d6804f9ed19f858a04b230f4250/fig13/eddy_streamline_outflow_flux_yplus_039.html}{(a)}, 
\href{https://cocalc.com/share/public_paths/cdee9e3cf9ce3d6804f9ed19f858a04b230f4250/fig13/eddy_streamline_outflow_flux_yplus_053.html}{(b)},
\href{https://cocalc.com/share/public_paths/cdee9e3cf9ce3d6804f9ed19f858a04b230f4250/fig13/eddy_streamline_outflow_flux_yplus_093.html}{(c)}, 
\href{https://cocalc.com/share/public_paths/cdee9e3cf9ce3d6804f9ed19f858a04b230f4250/fig13/eddy_streamline_outflow_flux_yplus_197.html}{(d)},
\href{https://cocalc.com/share/public_paths/cdee9e3cf9ce3d6804f9ed19f858a04b230f4250/fig13/eddy_streamline_outflow_flux_yplus_298.html}{(e)},
and for the inflow event at 
\href{https://cocalc.com/share/public_paths/cdee9e3cf9ce3d6804f9ed19f858a04b230f4250/fig13/eddy_streamline_inflow_flux_yplus_039.html}{(f)},
\href{https://cocalc.com/share/public_paths/cdee9e3cf9ce3d6804f9ed19f858a04b230f4250/fig13/eddy_streamline_inflow_flux_yplus_053.html}{(g)},
\href{https://cocalc.com/share/public_paths/cdee9e3cf9ce3d6804f9ed19f858a04b230f4250/fig13/eddy_streamline_inflow_flux_yplus_093.html}{(h)},
\href{https://cocalc.com/share/public_paths/cdee9e3cf9ce3d6804f9ed19f858a04b230f4250/fig13/eddy_streamline_inflow_flux_yplus_197.html}{(i)},
\href{https://cocalc.com/share/public_paths/cdee9e3cf9ce3d6804f9ed19f858a04b230f4250/fig13/eddy_streamline_inflow_flux_yplus_298.html}{(j)}.
 The code to generate outflow eddies is available \href{https://cocalc.com/share/public_paths/cdee9e3cf9ce3d6804f9ed19f858a04b230f4250/fig13/plot_outflow_flux_eddies.ipynb}{here} and to generate inflow eddies is available \href{https://cocalc.com/share/public_paths/cdee9e3cf9ce3d6804f9ed19f858a04b230f4250/fig13/plot_inflow_flux_eddies.ipynb}{here}.
}}
\label{fig:_eddy_flux}
\end{figure}}

{The difference in vorticity transport contributions between ``hairpins'' and ``shawls'' 
can be further emphasized by plotting as functions of $y$ the conditional mean vorticity flux 
$\langle v\omega_z-w\omega_y\rangle_{\pm,y_c}$ for both signs $\pm$ of wall-normal velocity and 
for various values of $y_c$ in the log-layer. See Figure~\ref{num}. Most obviously, the flux contribution
at the conditioning point $y=y_c$ of the $v_+$ eddies is always down-gradient, whereas the contribution 
of the $v_-$ eddies at point $y=y_c$ is always up-gradient. Another important lesson to draw from 
the flux distributions in Fig.~\ref{num} is the locality of the vorticity transport in vertical 
height, with eddies at wall distance $y_c$ contributing to vorticity flux only at distances $y\sim y_c.$
In fact, for all values of $y_c$ and both signs $\pm,$ the mean vorticity transport arising from 
the conditional eddy appears only in the scale range $0.4\lesssim y/y_c\lesssim 1.4$ and with 
a narrower range for larger $y_c.$ This result is complementary to the locality in spanwise 
length observed by \cite{kumar2023flux}, whose plots of vorticity-flux cospectra in their 
Figures 8 \& 11 showed that down-gradient vorticity across height $y$ is contributed on average 
by eddies with spanwise wavelengths $\lambda_z$ in the range $0.4\lesssim \lambda_z/y\lesssim 3$
while up-gradient vorticity across $y$ is contributed on average by eddies with $\lambda_z$ 
in the range $3\lesssim \lambda_z/y\lesssim 40.$ These two forms of locality are presumably closely
connected, since our Fig.~\ref{boxsize} shows that all three dimensions of the conditional mean eddies 
scale with $y_c.$ The observation in \cite{kumar2023flux} that up-gradient transport 
at wall distance $y$ arises from eddies of larger scale while down-gradient transport 
arises from eddies of smaller scale is also suggested by our Fig.~\ref{num}, since 
(at least for $y_c^+\lesssim 93$) the peak of the up-gradient transport from $v_-$ eddies
occurs at $y<y_c$ and the peak of the down-gradient transport from $v_+$ eddies occurs at $y>y_c.$ 
These various observations {\it in toto} lend support to Lighthill's conjecture that 
vorticity transport in wall-bounded turbulence occurs via a stepwise cascade through 
a hierarchy of eddies.}

\subsection{{Are Sweeps and Ejections Really Independent?}}

{Despite the differences documented above,}{ comparison of the inflow and outflow structures in Figure \ref{area} 
might lead 
one to question whether these two conditional mean eddies exist as distinct entities. Instead, the 
logarithmic layer at every wall distance may be imagined to consist of a sequence of streamwise vortices with 
alternating orientations, and the decision to pair these into ``hairpins'' and ``shawls'' would then 
be an arbitrary choice. It seems clear by continuity of the flow that outflows and inflows  
cannot be distributed independently but instead must appear one after another in turn, along a spanwise direction. In fact, this sort of arrangement corresponds exactly to the sketch in Fig.II.22 of \cite{lighthill1963}.} {Similarly in the plot of vortex lines in Fig.~\ref{fig:instant} for individual flow realizations,
one might imagine that the hairpin and inverted hairpin are two pieces of a long vortex line undulating up and down
in the spanwise direction and just shifted by half a phase.}

{To address these questions, we have carried out an extensive study of the distribution of inflow and outflow events 
identified for conditional averaging, with details given in Appendix \S \ref{sec:dist_pdf}.
(We thank an anonymous referee for suggesting this analysis.) 
Given any sweep or ejection event, we located the mean position of the neighboring reverse flow event by looking at the autocorrelation function of the wall normal velocity, in the same wall-normal plane. At each wall-height, we observe a clear negative minimum of the auto-correlation at spanwise distance of $\sin ±1.3y$, which we take as the average distance to the neighboring reverse flow event. This seems a reasonable estimate also based on the conditional mean structures plotted in Fig.10 and the individual realizations plotted in Fig.~\ref{fig:instant} (see also Figs.19 \& 20). However, these neighboring reverse events might not be sufficiently strong to satisfy the criterion imposed in our conditional ensemble. Thus, for each strong outflow event or ejection considered in the conditional average, we find the nearest strong inflow event or sweep on the same wall-parallel plane satisfying $|v|>v_{rms}$. We then find the displacement vector ${\bf d}$ between them. We consider displacement vectors ${\bf d}$ that make an angle 
$|\theta|<\frac{\pi}{4}$ with the $z$-axis and label these sweep events as possible ``spanwise shifts'' of the ejection event.
In fact, we find that these nearest sweep events are either much further away than $1.3y$ or else are displaced mainly streamwise, i.e. make an angle $|\theta|>\frac{\pi}{4}.$  At increasing wall normal distances in the log-layer, only 9\%-26\% sweep events are shifts of ejection events defined in this way. This result implies that, although a spanwise phase shift from a strong outflow event may lead to an inflow event, the strength of this inflow event is usually not enough to merit inclusion in the conditional average, and vice versa. Strong outflow and strong inflow events tend not to be phase shifts of each other, but instead perceptibly different events.}

{{In addition}, the conditional eddies for outflows and inflows are distinct in several 
respects. For example, the ``hairpins'' are more streamwise extended (see Fig.~\ref{boxsize}
for a quantification of this observation) and seem also more clearly ``wall-attached'', whereas 
the ``shawls''are streamwise shorter and appear detached. Even clearer differences appear in their 
contributions to vorticity transport. This becomes apparent in Figure~\ref{fig:_eddy_flux} 
where the conditional mean eddies are colored instead by nonlinear vorticity flux. ``Hairpins'' and 
``shawls'' both contain vorticity flux directed away from and toward the wall, but the distributions 
are distinctly different. For ``hairpins'' the dominant flux is outward/down-gradient and appears on the 
upper half of the vortex, whereas weaker inward/up-gradient flux appears on the underside facing the wall.
For ``shawls'' the distribution is different, with the dominant inward/up-gradient flux 
appearing on top and immediately below, with weaker outward/down-gradient flux further underneath. 
The most distinctive and important 
difference is that the arch of spanwise vorticity in ``hairpins'' is moving outward and contributes 
down-gradient vorticity flux, while the arch of spanwise vorticity in ``shawls'' is moving inward 
and contributes up-gradient vorticity flux. In fact, there is a striking similarity to the coherent 
structures observed by \cite{kumar2023flux} in filtered fields, designed to decompose the flow 
into two orthogonal components contributing ``down-gradient'' and ``up-gradient'' transport. 
Fig.13 in \cite{kumar2023flux} for the high-pass
filtered field shows a forest of ``hairpins'' with the same bipolar flux distribution as in 
the upper row of Fig.~\ref{fig:_eddy_flux} and net ``down-gradient''
transport. On the other hand, Fig.15 in \cite{kumar2023flux} for the low-pass filtered 
field shows an assembly of ``shawls'' or ``pancakes'' with the bipolar flux distribution as in 
the bottom row of Fig.~\ref{fig:_eddy_flux} and net ``up-gradient'' transport. The suggestive similarities 
between the structures revealed by conditional averaging and by spectral filtering remain to be 
fully understood.} 

{Further evidence that ``ejections'' and ``sweeps'' are not just phase shifts of each other is provided by the distribution of the nonlinear vorticity flux in individual realization of the two conditional ensembles, which are exhibited as colorplots in Figs. 19 \& 20 in Appendix \ref{sec:lines}. As suggested by the conditional mean structures plotted in Fig.~\ref{fig:_eddy_flux}, ejection events tend to have strong down-gradient flux just above the conditioning point while sweeps tend instead to have strong up-gradient flux just below that point.The neighboring reverse flow events are usually weaker and do not have these characteristics. These features are most evident nearer the wall where the individual realizations are more coherent, but remain as a statistical tendency at all wall distances. These differences in vorticity flux distributions 
for individual realizations of the two conditional ensembles give additional support to the premise that they 
represent different types of events, which are not just spanwise phase-shifts of each other.}

\iffalse
QUESTIONS:

\vspace{10pt}\noindent 
HAVE WE REALLY ``EXPLAINED" THE ANTI-CORRELATION IN FIGURE 1? WE SEE ANTI-CORRELATION ONLY BELOW THE CONDITIONING POINT FOR $v>0$ IN FIGURE 7, ONLY ABOVE THE CONDITIONING POINT FOR $v<0$ IN FIGURE 8.

DOUBTS:
\begin{eqnarray*}
&& \langle v(x,y,z)\omega_z(x,y,x)| v(x,y,z)>0\rangle \cr
&& \neq \langle v(x,y,z)| v(x,y,z)>0\rangle \times \left.\langle \omega_z(x',y',z')| v(x,y,z)>0\rangle \right|_{(x',y',z')=(x,y,z)}
\end{eqnarray*} 
and 
\begin{eqnarray*}
&& \langle w(x,y,z)\omega_y(x,y,z)| v(x,y,z)>0\rangle \cr
&& \neq \langle w(x,y,z)| v(x,y,z)>0\rangle \times \langle \omega_y(x',y',z')| v(x,y,z)>0\rangle _{(x',y',z')=(x,y,z)}
\end{eqnarray*} 

SUGGESTION: COMPARE

\begin{eqnarray*}
&& \langle v(x',y',z')\omega_z(x',y',z')| v(x,y,z)>0\rangle \qquad {\rm AND} \cr
&& \langle v(x',y',z')| v(x,y,z)>0\rangle \times \langle \omega_z(x',y',z')| v(x,y,z)>0\rangle 
\end{eqnarray*} 

ALSO 

\begin{eqnarray*}
&& \langle w(x',y',z')\omega_y(x',y',z')| v(x,y,z)>0\rangle \qquad {\rm AND} \cr 
&& \langle w(x',y',z')| v(x,y,z)>0\rangle \times \langle \omega_y(x',y',z')| v(x,y,z)>0\rangle
\end{eqnarray*} 
\fi 
\section{Conclusions}

The mechanism proposed by \cite{lighthill1963} for concentration of spanwise vorticity 
at solid walls in a turbulent boundary layer involves a strong correlation between
vortex-stretching/relaxation and fluctuating velocities toward/away from the wall. 
The method of conditional averaging \citep{kim1986structure,adrian1989}
is designed to reveal such correlations
and, applied here to a database of high-$Re_\tau$ turbulent channel flow, it provides extensive evidence corroborating the validity of Lighthill's mechanism throughout the logarithmic layer. 
We have elaborated this picture by observing, in addition to the near-wall motions postulated by \cite{lighthill1963}, also returning counter-flows away from the wall, which help to explain,  among other things, the anti-correlation between advective and stretching fluxes observed by \cite{kumar2023flux}. The net vorticity transport in the turbulent boundary layer is exposed as an intense rivalry between fluxes up-gradient toward the wall and down-gradient away from the wall, with the competition just narrowly won by the latter.  Our results further support 
{by an array of evidence} 
Lighthill's conjecture that this vorticity dynamics is a cascade process proceeding through a hierarchy of turbulent eddies whose dimensions scale with distance to the wall. 

The present work supports the view that the current AEM could be improved by inclusion of up-gradient transport. {\cite{eyink2008} pointed out that the 
nonlinear vorticity flux obtained by matched asymptotics over the entire log-layer 
is reproduced within the AEM as 
\be \langle v\omega_z-w\omega_y\rangle\propto u_\tau^2\left(\frac{P\delta}{y^2}-\frac{Q}{h}\right) \ee
for $\delta:=\nu/u_\tau\ll y\ll h,$ if only one replaces the usual AEM assumption on the 
eddy-intensity function that $I_{xy}(y^*)\to 0$ for $y^*\gtrsim 1$ instead with $I_{xy}(y^*)\sim 
-P/y^*$ for $y^*\gg 1.$ In that case, up-gradient vorticity transport for $y<y_p:=(P\delta h/Q)^{1/2}$ 
is correctly recovered, but the assumed decay is much slower than the rate $I_{xy}(y^*)=O(1/y^{*4})$ for $y^*\gg 1$ obtained from the Biot-Savart formula by assuming a ``representative eddy" in the form of a hairpin line-vortex \citep[][Appendix C]{woodcock2015statistical}. 
It has been an open question since \cite{eyink2008} what alternate vortex structure 
might yield such up-gradient transport. In agreement with the classic work of \cite{kim1986structure},
our study suggests that the additional structures could be associated with sweeps and we identify 
these as ``shawl vortices''.}  
Lighthill's hypothesis that vorticity transport toward the wall is taken over by 
``smaller-scale movements'' suggests another competitive direct-cascade process in which large-scale vortices are transported toward the wall, fragment into smaller structures and proliferate in number. The shawl vortices wrapped around downflows, as revealed by conditional averaging in Fig.~\ref{eddy}(b), seem to 
originate above the wall but interact with it by inducing opposite-sign vorticity. These observations
open up new possibilities for vortex-based structure models that may refine the 
attached-eddy model. 

There are many possible elaborations and future directions of work. 
\iffalse 
It is unknown
how well the linear estimator approximates the exact conditional mean. 
The conditional eddies exploited for our fluid mechanical explanations furthermore underestimate the Lighthill effect of the stretching term below the conditioning point and overestimate its 
down-gradient effect above, because of statistical correlations in magnitudes not yet fully understood. 
\fi 
We have not fully explored the 3D characteristics of the conditional eddies illustrated in Fig.~\ref{eddy}. For example, the control-volume 
arguments presented in Figs.~\ref{outflow_sketch},\ref{inflow_sketch} are essentially 
2D but the vortex lines plotted in Fig.~\ref{eddy}(a),(b) are streamwise inclined 
and likewise the conditional flows are fully 3D. 
{Perhaps most importantly, our study supports Lighthill's suggestion of a vorticity cascade mechanism, but no strict causal connection has been established between vorticity at different 
scales, locations, and times}. A promising method to get more 
detailed dynamical understanding is the stochastic Lagrangian approach \citep{wang2022origin},
especially if combined with conditional averaging. Monte Carlo evaluation of the stochastic 
Lagrangian trajectories is prohibitively expensive in the logarithmic layer, but a 
recent Eulerian adjoint vorticity algorithm \citep{xiang2025origin} makes this feasible. 
{In the latter paper, Lighthill's mechanism was verified as the causal origin 
of strong spanwise vorticity in the viscous sublayer of turbulent channel flow.} 
Finally, an important direction of {future} research is to exploit better understanding of the vorticity dynamics responsible for turbulent drag in order to develop improved drag reduction strategies \citep{kumar2025josephson}. 

%\noindent{\bf Acknowledgements.} We thank the JHTDB team for their help and support of the public numerical turbulence laboratory.

\vskip 0.1in

\noindent{\bf Funding:}
We thank the Simons Foundation for support of this work through the Targeted Grant No. MPS-663054 and Collaboration Grant No. MPS-1151713 
and the National Science Foundation for funding through Grant \# CSSI-2103874.

\vskip 0.1in

\noindent{\bf Declaration of interests.}  The authors report no conflict of interest.

\vskip 0.1in

\noindent{\bf Author ORCIDs.} \\
S. Kumar \href{https://orcid.org/0000-0002-6785-0072}{https://orcid.org/0000-0002-6785-0072}.\\
S. Toedtli \href{https://orcid.org/0000-0001-9371-9572}{https://orcid.org/0000-0001-9371-9572}.\\
T. Zaki \href{https://orcid.org/0000-0002-1979-7748}{https://orcid.org/0000-0002-1979-7748}. \\
G. Eyink \href{https://orcid.org/0000-0002-8656-7512}{https://orcid.org/0000-0002-8656-7512}.

\iffalse
\vspace{-15pt} 
\bibliographystyle{jfm}
\bibliography{jfm}
\newpage 
\renewcommand{\thesection}{\Alph{section}}
\setcounter{section}{0} 
\setcounter{page}{1} 
\renewcommand\thefigure{\thesection.\arabic{figure}}    
\setcounter{figure}{0}    
\fi 
%\begin{center} 
%{\large {\bf SUPPLEMENTARY MATERIALS}} 
%\end{center}

%\tableofcontents 
\appendix 
%\newpage
%\clearpage 

\clearpage 

\section{Area and Flux Fractions of High Wall-normal Velocity Points}\label{sec:frac} 

\begin{figure}
\begin{center}
{\includegraphics[width=\textwidth]{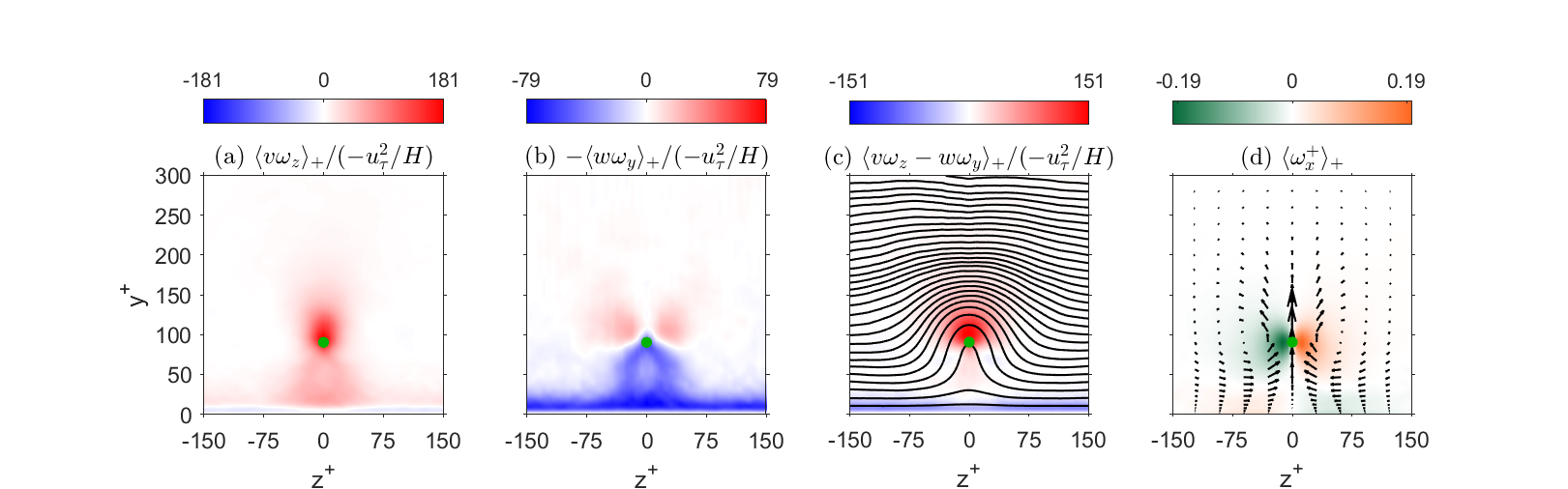}}
\end{center}
\caption{Conditional mean fields {in the plane of the conditioning point} for the outflow event $v>2v_{rms}$ at $y_c^+=92.8$, colored by (a) flux due to the convective term, (b) flux from the stretching/tilting term, (c) total nonlinear flux, {(d)} streamwise vorticity. \iffalse Colorbars are asymmetric in (a)-(c) to emphasize sign changes. \fi Also depicted are (c) vortex lines and (d) quivers showing in-plane velocity. A green dot marks the conditioning point.}
\label{outflow_2rms} 
\end{figure}

\begin{figure}
\begin{center}
{\includegraphics[width=\textwidth]{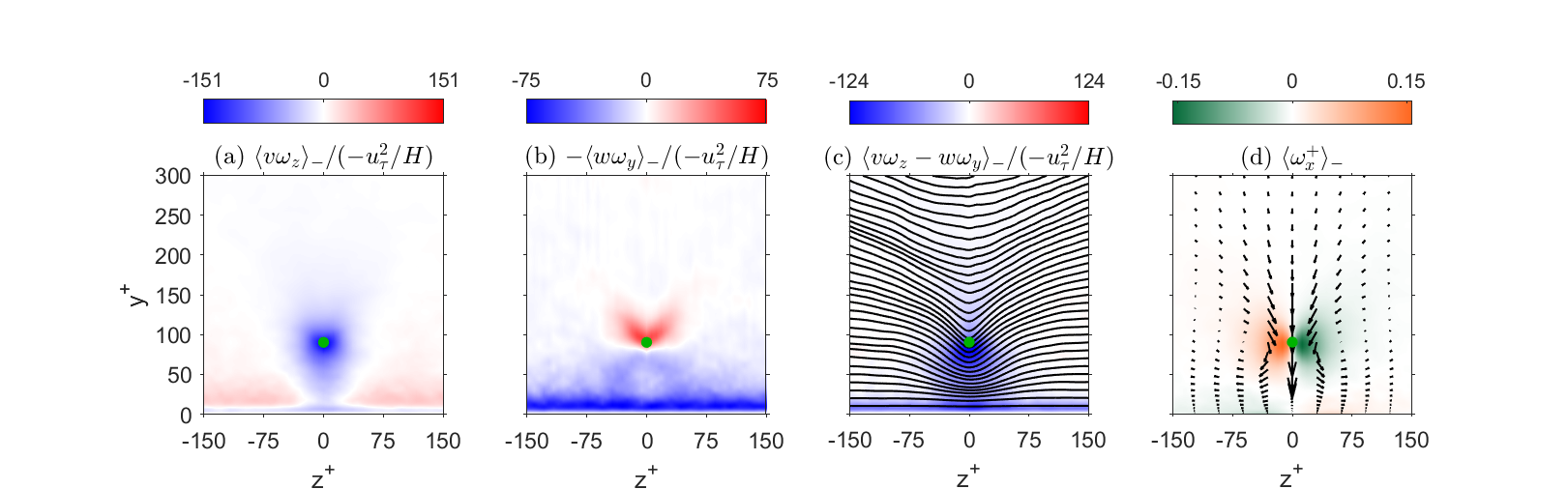}}
\end{center}
\caption{{Conditional mean fields {in the plane of the conditioning point} for the inflow event $v<-2v_{rms}$ at $y_c^+=92.8$, colored by (a) flux due to the convective term, (b) flux from the stretching/tilting term, (c) total nonlinear flux, {(d)} streamwise vorticity. Also depicted are (c) vortex lines and (d) quivers showing in-plane velocity. A green dot marks the conditioning point.}}
\label{inflow_2rms} 
\end{figure}

{The average fluxes displayed in Fig.~\ref{fig:fluxes} in the main text were conditioned 
simply on $v>0$ and $v<0$, but the definition of the conditional ensembles in section~\ref{sec:num} 
employed a threshold magnitude of wall-normal velocity. To select an appropriate threshold, we applied 
conditions $v>\alpha v_{rms}$ and $v<-\alpha v_{rms}$ for various choices of the parameter $\alpha.$ 
For each choice, we calculated the fractional area occupied by points satisfying that condition 
in wall-parallel planes at fixed heights $y_c$ and also the fractional contribution of those points 
to the mean nonlinear vorticity flux through those planes.}

\begin{figure*}
 \centering
{\includegraphics[width=0.8\textwidth]{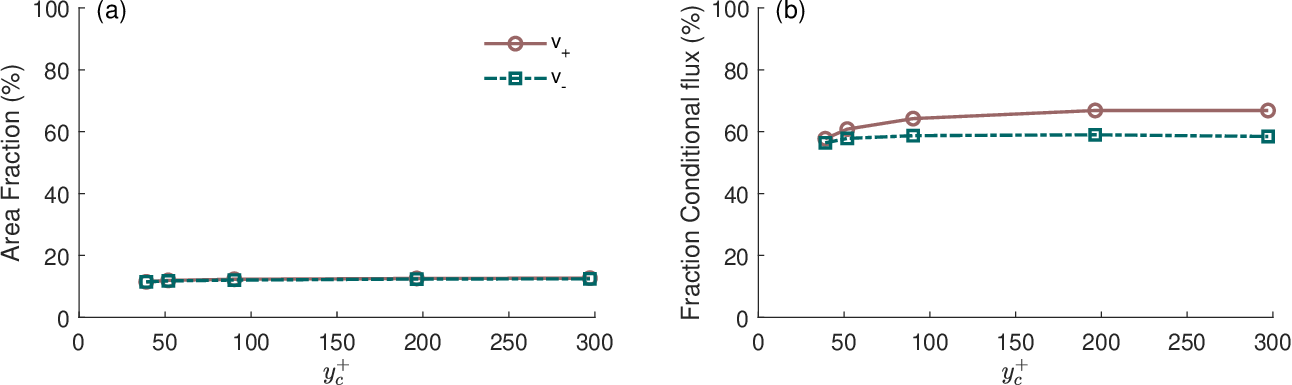}}
\caption{{(a) Fraction of area occupied, and (b) contribution to conditional flux from strong outflow events($v>v_{rms}$) and strong inflow events ($v<v_{rms}$) }  }
\label{fig:fluxes_cond} 
\end{figure*}

{For the case $\alpha=1$ shown in Fig.~\ref{fig:fluxes_cond} we found that the points 
for both $v_+$ and $v_-$ conditions occupy around 12\% of the area, nearly independent of $y_c.$
On the other hand, the contribution of these points to the nonlinear vorticity flux 
through those planes is upward of 60\% and very slightly increasing with $y_c$. 
Thus, the flux contribution outweighs the area fraction by more than fivefold, emphasizing 
the importance of these regions for vorticity transport. We note also that at each $y_c$-value 
there are somewhat more points satisfying the $v_+$ condition than the $v_-$ condition, with 
the difference growing slightly as well with increase of $y_c$. This observation
is another indication of the somewhat greater strength of outflow events than inflow events 
at all wall distances.}  

\begin{figure*}
 \centering
{\includegraphics[width=0.8\textwidth]{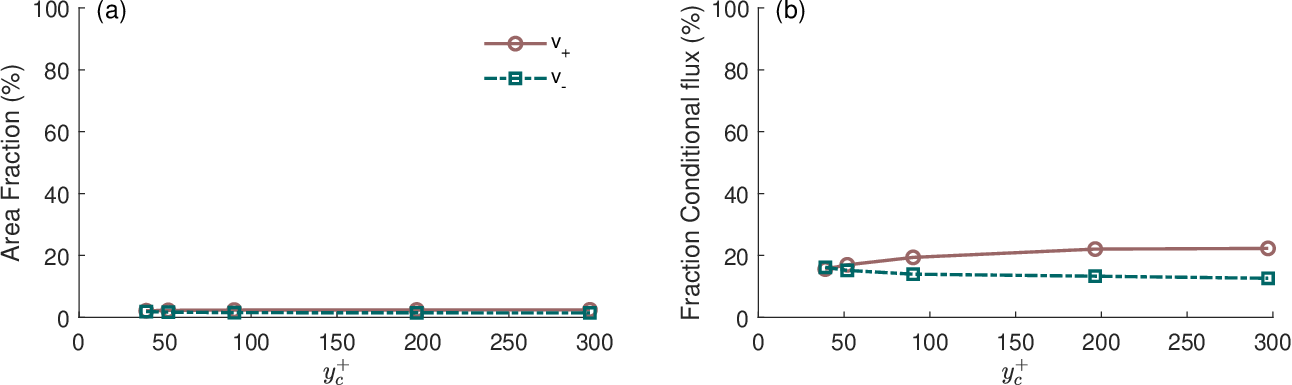}}
\caption{{(a) Fraction of area occupied, and (b) contribution to conditional flux from strong outflow events($v>2v_{rms}$) and strong inflow events ($v<2v_{rms}$) }  }
\label{fig:fluxes_cond_2rms} 
\end{figure*}

{For the case $\alpha=2$ shown in Fig.~\ref{fig:fluxes_cond_2rms} we found that the points 
for both $v_+$ and $v_-$ conditions occupy around 2\% of the area, again nearly independent of $y_c,$
but the contribution to the nonlinear vorticity flux is upward of 20\% and 
again slightly increasing with $y_c$. Thus, the flux contribution outweighs the area fraction 
here by more than tenfold. We observe once again greater strength of outflow events than inflow 
events at all wall distances. Furthermore, we calculated the conditional means fields 
for $\alpha=2$ analogous to those plotted in Figs.~\ref{outflow},\ref{inflow}, using the 
same methodology described in section \ref{sec:num}. The results are extremely close to those 
presented in Figs.~\ref{outflow},\ref{inflow} of the main tex for $\alpha=1,$ with just somewhat 
greater magnitudes of all conditional mean fields.}  

{Based on all of these observations, we decided to present in the main body of the paper 
results for the case $\alpha=1$. As we have seen here, this choice of threshold was sufficiently
high to obtain coherent flow structures from conditioning and, at the same time, sufficiently low 
to guarantee nearly complete coverage of the wall area by the sampling windows for the selected events.}

\begin{figure*}
\centering
{\includegraphics[width=\textwidth]{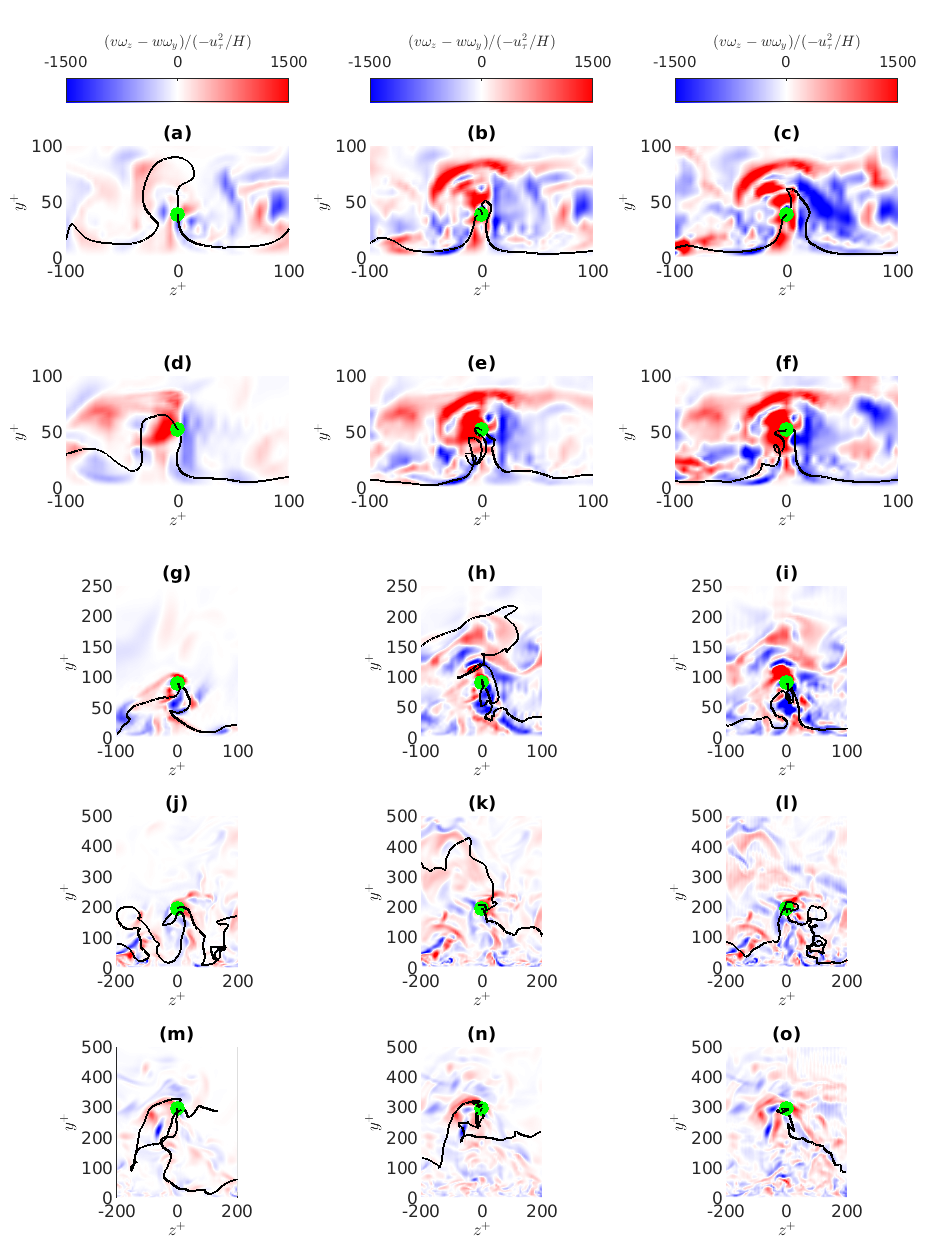}}
\caption{ {Vortex lines passing through the conditioning point associated with the strongest outflow events at a given wall height, at different time instants,{with the background colored by the instantaneous nonlinear flux on the spanwise-wall normal plane passing through the conditioning point.} The conditioning points are at (a-c)$y_c^+=39$, (d-f) $y_c^+=52$, (g-i)$y_c^+=92.8$, (j-l)$y_c^+=197$, (m-o) $y_c^+=298$. }}
\label{lines-out}
\end{figure*}

\begin{figure*}
\centering
{\includegraphics[width=\textwidth]{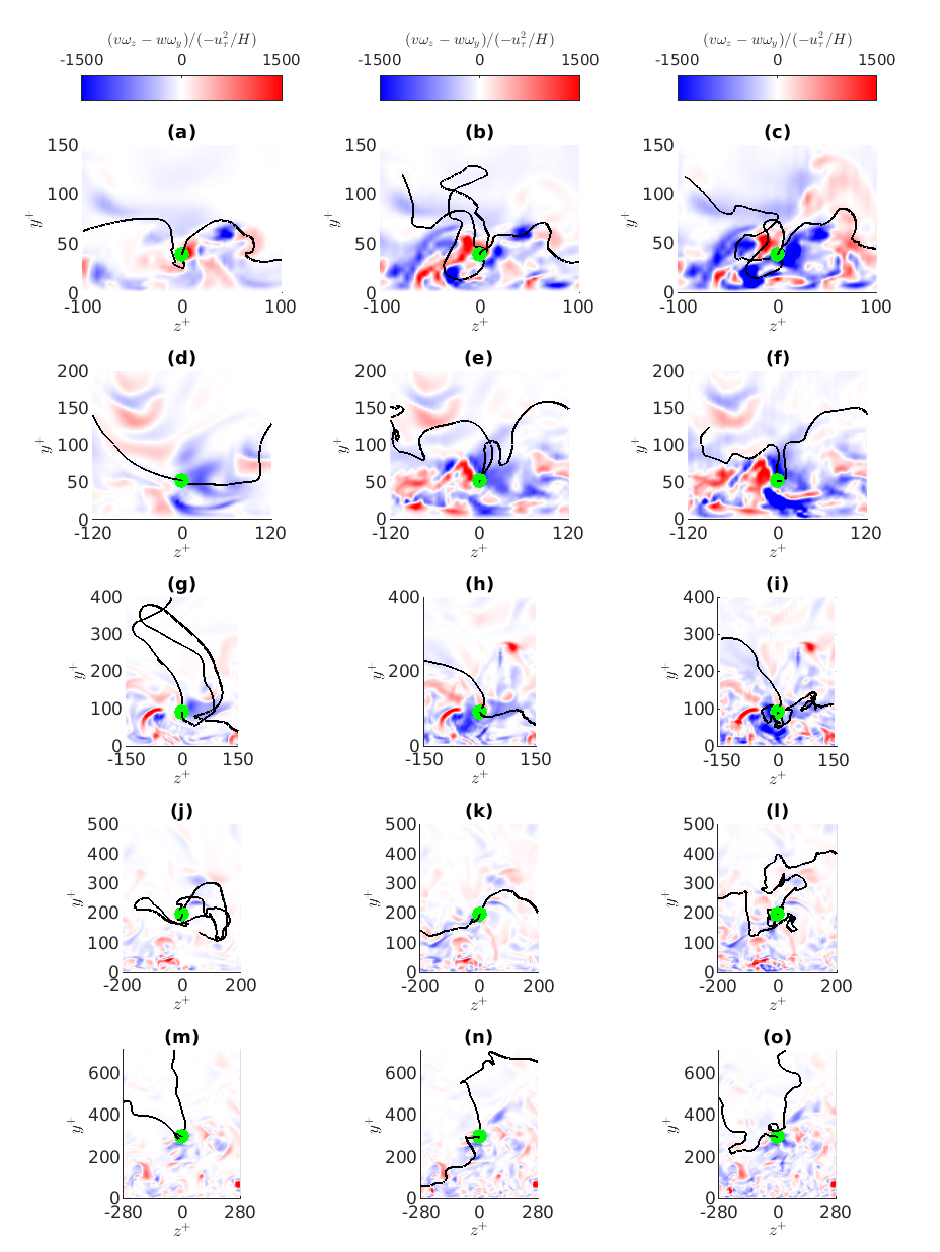}}
\caption{{Vortex lines passing through the conditioning point associated with the strongest inflow events at a given wall height, at different time instants,{with the background colored by the instantaneous nonlinear flux on the spanwise-wall normal plane passing through the conditioning point.} The conditioning points are at (a-c)$y_c^+=39$, (d-f) $y_c^+=52$, (g-i)$y_c^+=92.8$, (j-l)$y_c^+=197$, (m-o) $y_c^+=298$.}}
\label{lines-in}
\end{figure*}

\section{Vortex Lines in Individual Realizations of the Conditional Ensembles}\label{sec:lines}

%\fi
%\clearpage 
%\newpage 

%\newpage 
%\newpage
%\clearpage 

{At the end of section \ref{sec:num}, we exhibited vortex lines for the strongest 
outflow ($v_+$) and inflow ($v_-$) events, as identified by our conditional sampling method in one time snapshot at $y_c^+=92.8.$ To help give more intuition about the individual realizations 
of our conditional ensembles, we here plot vortex lines for additional snapshots. 
Furthermore, to see the variation with wall distance, we show vortex lines
in Fig.~\ref{lines-out} for the strongest outflow events at five wall distances
distributed through the log-layer in three time snapshots and in Fig.~\ref{lines-in} 
we show vortex lines 
for the strongest inflow events in the same three snapshots at the same wall distances. 
To keep the plots simple, we show the single unique vortex line that passes through 
the conditioning point and projected into the $yz$-plane through that point.} 

{Although vortex lines are plotted for only a few events out of the large 
number in the conditional ensembles (see Table \ref{tab:num_events}), 
they suggest a few general trends.
First, we see that outflow events lead to hairpin-type vortices but more or less 
disordered by the turbulent environment and likewise the inflow events lead to 
inverted hairpin-type vortices. The vortex lines are more ordered for outflows 
than for inflows and also more ordered at decreasing distances from the wall.} 

{The vortex lines through the conditioning points, as well as those at a fraction 
of the wall-distance higher and lower, can be reconstructed from the earlier-in-time 
vorticity in a Lagrangian sense by the methods of \cite{wang2022origin,xiang2025origin}.
Such reconstruction for the conditional ensembles can reveal the causal 
dynamics of the vorticity cascade proposed by \cite{lighthill1963}.}
%\clearpage 

\begin{figure*}
\centering
{\includegraphics[width=\textwidth]{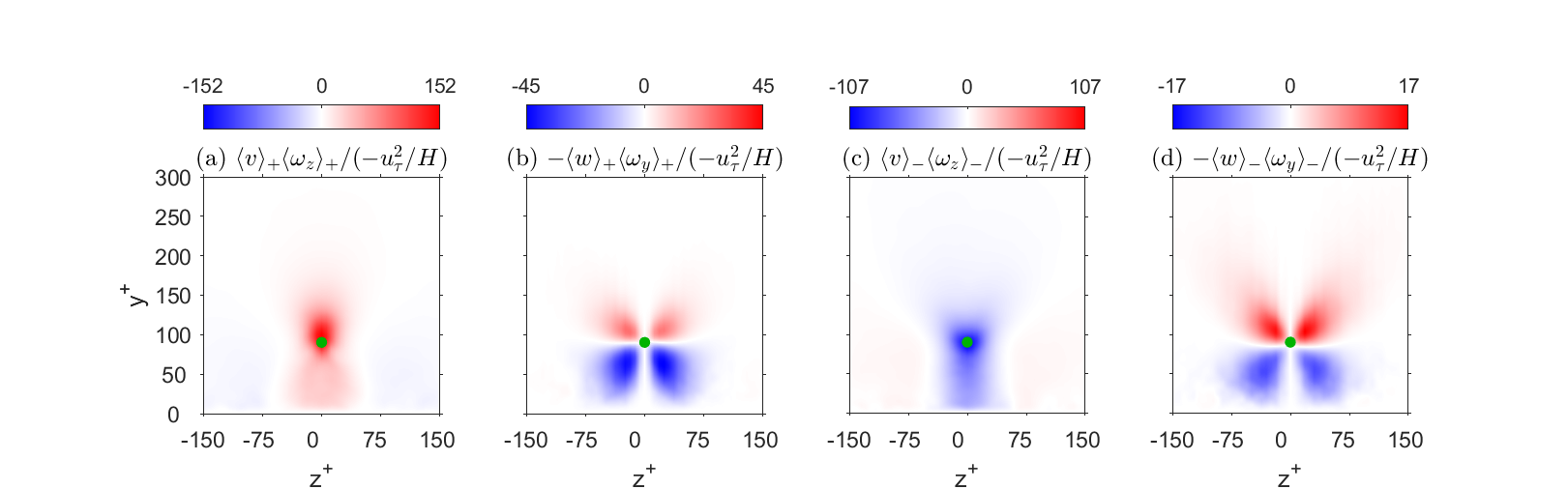}}
\caption{Vorticity flux fields of mean eddies conditioned on the outflow/inflow events at $y_c^+=92.8:$ 
the convective term for (a) outflow and (c) inflow and the stretching/tilting term for (b) outflow, and (d)
 inflow. A green dot marks the conditioning point.}
 %A dashed curve marks a change in sign.}
\label{fluxcond} 
\end{figure*}

\begin{figure*}
\centering
{\includegraphics[width=\textwidth]{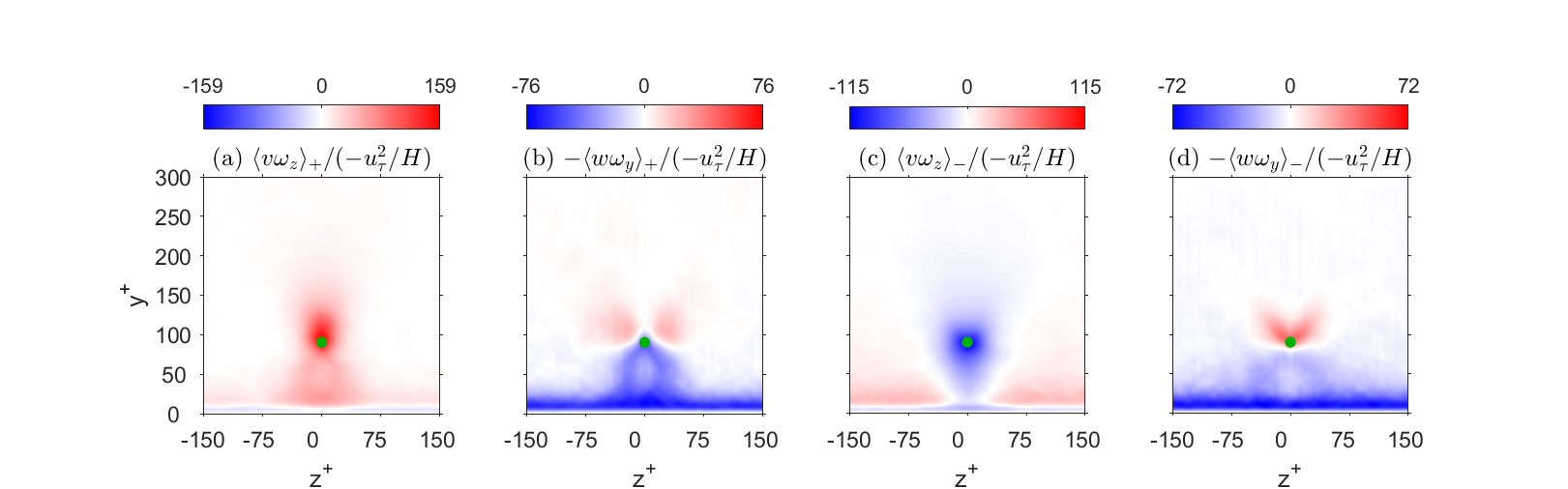}}
\caption{Mean vorticity flux fields conditioned on the outflow/inflow events at $y_c^+=92.8:$
the convective term for (a) outflow and (c) inflow and stretching/tilting term for (b) outflow
and (d) inflow. A green dot marks the conditioning point.} 
%A dashed curve marks a change in sign.}
\label{condflux} 
\end{figure*}

\begin{figure*}
\centering
{\includegraphics[width=0.6\textwidth]{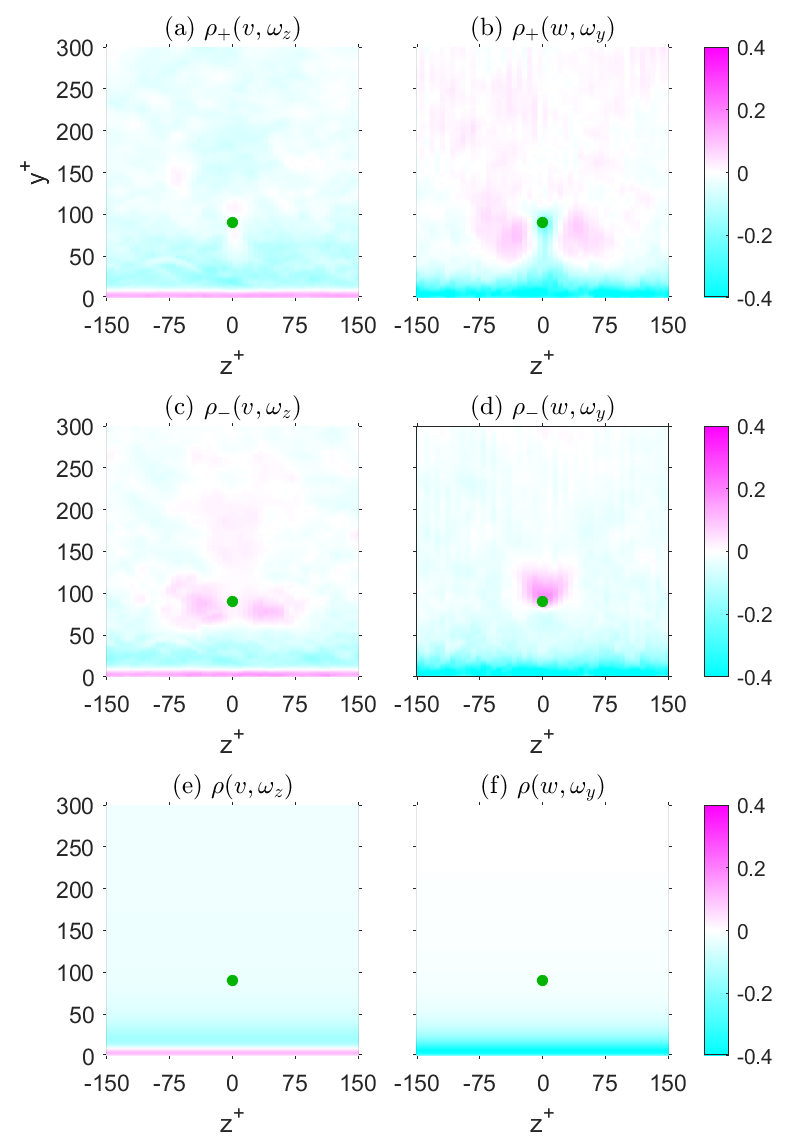}}
\caption{Correlation coefficients for the two factors in the convective term $\rho_{\pm}(v,\omega_z) = (\langle v\omega_z \rangle_{\pm} - \langle v\rangle_{\pm} \langle \omega_z\rangle_{\pm})/v^{rms}_\pm\omega_{z\pm}^{rms},$ conditioned on (a) outflow, and (c) inflow, 
and for the two factors in the stretching/tilting term, $\rho_{\pm}(w,\omega_y) = (\langle w\omega_y \rangle_{\pm} - \langle w\rangle_{\pm} \langle \omega_y\rangle_{\pm})/w^{rms}_\pm\omega_{y\pm}^{rms},$ 
conditioned on (b) outflow and (d) inflow, at $y_c^+=92.8$. Also shown are the unconditioned correlation coefficients for the convective term (e)$\rho(v,\omega_z) = (\langle v\omega_z \rangle - \langle v\rangle \langle \omega_z\rangle)/v^{rms} \omega_{z}^{rms}$ and stretching/tilting term (f)$\rho(w,\omega_y) = (\langle w\omega_y \rangle - \langle w\rangle \langle \omega_y\rangle)/w^{rms}\omega_{y}^{rms}$ .} %NORMALIZED HOW???}
\label{Pcorr} 
\end{figure*}

\section{Negligible Effect of Fluctuations in Conditional Mean Fluxes}\label{sec:corr} 

{We noted in the main text at the end of section \ref{sec:in} that our 
successful explanation
of the conditional mean vorticity flux based upon the properties of the conditional mean eddies 
requires that correlations of velocity and vorticity fluctuations be weak. Here we expand on that point 
in some detail and we furthermore quantify the magnitude of velocity-vorticity 
fluctuation correlations}.

{
In the first Figure~\ref{fluxcond} we plot the fluxes of the conditional eddies, 
both convective $\langle v\rangle_\pm\langle \omega_z\rangle_\pm$ and stretching/tilting 
$-\langle w\rangle_\pm\langle \omega_y\rangle_\pm,$ which are directly related 
to the conditional field lines and flows plotted in Fig.~\ref{outflow}(c),(d) and 
Fig.~\ref{inflow}(c),(d). These quantities are {\it a priori} distinct from the conditional mean fluxes, 
$\langle v\omega_z\rangle_\pm$ and $-\langle w \omega_y\rangle_\pm,$ plotted in 
Fig.~\ref{outflow}(a),(b) and Fig.~\ref{inflow}(a),(b) in the main text and 
reproduced here for convenience in Fig~\ref{condflux}. Since the issues are 
very similar for all $y_c$ values, we have confined the plots and discussion 
to the single wall-normal distance $y_c^+=92.8$ presented as the primary example 
in the main text.} 

{Comparing the results in Figs.~\ref{fluxcond},\ref{condflux}, we see first they 
are remarkably similar, except near the wall. Not only are the signs and patterns of the 
two sets of fluxes closely similar, but also the magnitudes are quite similar for the advective flux
and differ by a factor of only 2-3 for the stretching flux. On the other hand, the conditional mean 
fluxes show up-gradient transport very near the wall that is missing in the flux of the 
conditional eddies, especially strong for the stretching fluxes.  Furthermore, this near-wall 
up-gradient flux is quite uniform in $z$ and also nearly the same for outflow and inflow events.
It is reasonable to expect that memory of the condition at $y_c^+=92.8$ fades far from the 
conditioning point, so that the near-wall conditional mean fluxes are presumably close 
to the unconditional means.}

{The close resemblance of the results in Figs.~\ref{fluxcond},\ref{condflux} requires
that the fluctuation correlation coefficients of the velocity and vorticity factors in the fluxes 
must be small for both condition ensembles. In Fig.~\ref{Pcorr}(a-d) we have plotted 
the Pearson correlation coefficients for both advective \& stretching/titling fluxes and
for both outflow/inflow. Except near the walls, the coefficients are less in magnitude 
than about 0.2, confirming the small correlations of fluctuations. 
The coefficients are strongest close 
to the wall, especially for the stretching flux where they rise to a maximum magnitude 
of 0.4. We have plotted also the Pearson correlation coefficients for the unconditional
means in Fig.~\ref{Pcorr}(e),(f), verifying that they are nearly the same as for the 
analogous conditional means in Fig.~\ref{Pcorr}(a-d). In the main text, we proposed 
that the generally weak correlations of fluctuations are due to the scale separation expected 
for velocity and vorticity and this explanation is consistent with the increase of 
correlations in the near-wall region where the separation in scales disappears.}

\section{{Distribution of inflow and outflow events }}\label{sec:dist_pdf}

\begin{figure}
\begin{center}
{\includegraphics[width=\textwidth]{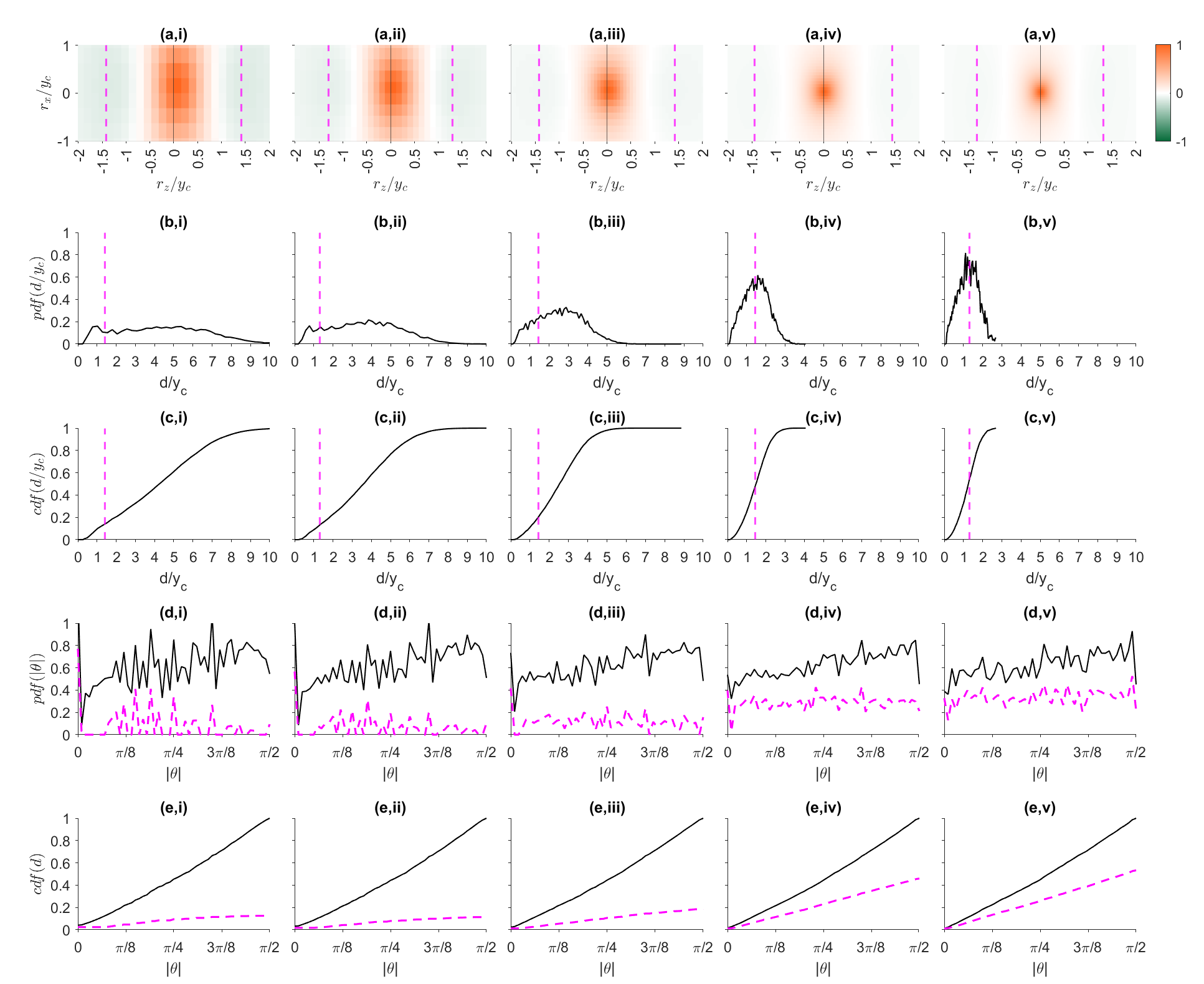}}
\end{center}
\caption{{(a,i-v)Autocorrelation of wall normal velocity $\rho_{vv}(r_x,y_c,r_z)$, magenta dashed lines mark the location of local minima, $r_z^+=r_{m}^+$ and $ r_z^+=- r_m^+$ . (b,i-v) Probability distribution function $(pdf)$ and (c,i-v) cumulative distribution function of the reverse event distance $d=|\mathbf{d}|$, with magenta dashed lines marking respective $r_m^+$. (d,i-v) Probability distribution function ($PDF$) and (e,i-v) cumulative distribution function ($CDF$) of the absolute value of the angle $|\theta|$ made by $\mathbf{d}$ with the $z$ axis. Magenta dashed lines show the $PDF$ and $CDF$ of $|\theta|$ where $d\leq r_m$. The planes are at (a-e,i)$y_c^+=39$, (a-e,ii)$y_c^+=52$, (a-e,iii)$y_c^+=92.8$, (a-e,iv)$y_c^+=197$, (a-e,v)$y_c^+=298.$  }}
\label{dist_pdf} 
\end{figure}

{The autocorrelation function of the normal velocity in wall-parallel planes at height $y_c$, 
$$\rho_{vv}(r_x,r_z|y_c)=\frac{\langle v(x,y_c,z,t) v(x+r_x,y_c,z+r_z,t) \rangle_{x,z,t}}{v_{rms}^2(y_c)},$$ shown in Fig~\ref{dist_pdf}(a,i-v), exhibits a weak negative mininum at spanwise distance $r_z=r_m\sim 1.3 y_c$ for each 
$y_c$ value. This minimum implies that on average, moving a distance $\pm r_m$ in spanwise direction from a strong outflow (inflow) event will lead us to an inflow (outflow) event.
However, the strength of the autocorrelation suggests that the inflow (outflow) event may be much weaker than the corresponding outflow (inflow) event.}

{On the other hand, the conditional ensembles that we have employed include only events where $|v|>v_{rms}$. A strong outflow (inflow) event included in the conditional average may or may not have an inflow (outflow) event which can be found by a spanwise shift of $r_m$ and which is also strong enough to be included in the corresponding conditional average. In order to quantify the proportion of such reverse events which are sufficiently strong, we define the reverse event displacement vector $\mathbf{d}$. At a given wall normal height $y_c$, for each outflow event included in the conditional average ($v(\mathbf{x}_p)>v_{rms}$) located at $\mathbf{x}_p=(x_p,y_c,z_p)$, let the nearest inflow event included in the conditional average ($v(\mathbf{x}_n)<-v_{rms})$) be located at $\mathbf{x}_n=(x_n,y_c,z_n)$. We then define $\mathbf{d}:=\mathbf{x}_n-\mathbf{x}_p=(x_n-x_p,0,z_n-z_p)$, with reverse event distance $d=|\mathbf{d}|$ and $\theta = \arctan\left( \frac{z_n-z_p}{x_n-x_p} \right)$, the angle made by the vector $\mathbf{d}$ with the $z$ axis. }
{We posit that an inflow event can be considered a ``spanwise shift'' of an outflow event if $d \leq r_m$ and $|\theta|\leq \pi/4$. The PDF and CDF of $d$ for different $y_c$ -values are shown in Fig.~\ref{dist_pdf}(b,i-v) and (c,i-v), with dashed pink lines marking the corresponding $r_m$. We see that as the wall normal distance $y_c$ of the conditioning point increases, the proportion of cases with $d<r_m$ also increases from $14\%$ for $y_c^+=39$ (Fig~\ref{dist_pdf}(c,i)) to $54\%$ for $y_c^+=298$(Fig~\ref{dist_pdf}(c,v). Meanwhile, the PDF and CDF of $\theta$ are illustrated in Fig.~\ref{dist_pdf}(d,i-v) and (e,i-v) respectively, with magenta dashed curves showing the corresponding values for the cases where $d \leq r_m$. The plots show that overall, there is no strongly preferred direction of displacement, while the proportion of cases which could be considered a ``spanwise shift'', with $|\theta|\leq \pi/4$ and $d \leq r_m$ increases from $9\%$ for $y_c^+=39$(Fig.~\ref{dist_pdf}(e,i)) to $26\%$ for $y_c^+=298$(Fig.~\ref{dist_pdf}(e,v)). }

{Thus, we see that at most a quarter of the inflow events strong enough to be included in the conditional average can be considered a ``spanwise shift'' of a strong outflow event. Our results show that strong outflow and strong inflow events cannot in general be identified with the adjacent reverse 
flows obtained from small spanwise shifts. This finding supports the idea that the ejections and sweeps identified 
by our criteria are not adjacent parts of the same structure, but instead different structures. It is true that small 
spanwise shifts from a strong outflow (inflow) event will lead to an inflow (outflow) event, but these adjacent reversed flows are generally 
weaker and do not satisfy the strength criterion that we impose to define our conditional ensembles.}

\bibliographystyle{jfm}
\bibliography{jfm}

%\newpage 
%\clearpage 

\end{document}